\documentclass[12pt,a4paper]{article}
\pdfoutput=1
\usepackage{jheppub}  
\usepackage{amssymb} 
\usepackage{amsmath}
\usepackage{mathtools}
\usepackage{amsfonts}    
\usepackage{dsfont}
\usepackage{pdfpages}
\usepackage{verbatim}
\usepackage{tensor}
\usepackage{mathrsfs}
\usepackage[mathscr]{euscript}
\usepackage{tikz}
\usetikzlibrary{hobby}

\newcommand{\ba}{\begin{align}}

\newcommand{\be}{\begin{equation}}
\newcommand{\ee}{\end{equation}}
\def\bd{\begin{tikzpicture}}
\def\ed{\end{tikzpicture}}
\newcommand{\ket}[1]{| #1\rangle}

\allowdisplaybreaks[1]

\title{Deriving the $\text{AdS}_{\bf 3}/\text{CFT}_{\bf 2}$ Correspondence}

\author[a,b]{Lorenz Eberhardt}
\author[a]{\!\!, Matthias R.\ Gaberdiel}
\author[c]{and Rajesh Gopakumar} 
\affiliation[a]{Institut f\"ur Theoretische Physik, ETH Zurich, \\
\hspace*{0.3cm}CH-8093 Z\"urich, Switzerland}
\affiliation[b]{School of Natural Sciences, Institute for Advanced Study, \\
\hspace*{0.3cm}Princeton, NJ 08540, USA}
\affiliation[c]{International Centre for Theoretical Sciences-TIFR, \\
\hspace*{0.3cm}Shivakote, Hesaraghatta Hobli, \\
\hspace*{0.3cm}Bengaluru North, India 560 089}
\emailAdd{elorenz@ias.edu}
\emailAdd{gaberdiel@itp.phys.ethz.ch}
\emailAdd{rajesh.gopakumar@icts.res.in}

\abstract{It was recently argued that string theory on 
${\rm AdS}_3\times {\rm S}^3\times \mathbb{T}^4$ with one unit ($k=1$) of NS-NS flux is exactly dual to the symmetric orbifold CFT  ${\rm Sym}^N(\mathbb{T}^4)$. In this paper we show how to directly relate the $n$-point correlators of the two sides to one another. 
In particular, we argue that the correlators of the world-sheet theory are delta-function-localised in string moduli space to those configurations that allow for a holomorphic covering map of the $\text{S}^2$-boundary of $\text{AdS}_3$ by the world-sheet. This striking feature can be seen both from a careful Ward identity analysis, as well as from semi-classically exact AdS$_3$ solutions that are pinned to the boundary. 
The world-sheet correlators therefore have exactly the same structure as in the Lunin-Mathur construction of symmetric orbifold CFT correlators in terms of a covering surface ---  which now gets identified with the world-sheet. Together with the results of \cite{Gaberdiel:2018rqv, Eberhardt:2018ouy} this essentially demonstrates how the $k=1$ $\text{AdS}_3$ string theory becomes equivalent to the spacetime orbifold CFT in the genus expansion.
}

\begin{document}

\maketitle

\makeatletter
\g@addto@macro\bfseries{\boldmath}
\makeatother

\section{Introduction} \label{sec:intro}

The AdS/CFT correspondence \cite{Maldacena:1997re} relates two autonomously defined mathematical \linebreak struc\-tures. Namely, perturbative string theory on ${\rm AdS}_{d+1}$ backgrounds (defined in terms of a worldsheet sigma model), and conventional quantum field theories at their conformal fixed points. One can therefore independently make computations on both sides and compare them using the dictionary that translates their observables to one another \cite{Gubser:1998bc, Witten:1998qj}. And indeed much effort has gone into this over twenty years and given strong evidence for the correspondence. However, one might ideally like to {\it derive} this correspondence by having a concrete way to transform observables on one side to the dual ones, manifesting thus their equality. Having such a transformation can give deeper insights into why this correspondence holds apart from obviating the need to check in a case by case manner.\footnote{There have been several attempts over the years to derive the AdS/CFT correspondence, or more generally, gauge-string duality at varying levels of specificity. A partial list of references include \cite{Gopakumar:1998ki, Ooguri:2002gx, Berkovits:2003pq, Gopakumar:2003ns, Gopakumar:2004qb, Gopakumar:2005fx, Gaiotto:2003yb, Berkovits:2007zk, Berkovits:2007rj, Berkovits:2008qc, Berkovits:2019ulm, Nastase:2018cfe}.} 

One context, where such a program can potentially be carried to completion, is string theory on ${\rm AdS}_3$ backgrounds with NS-NS flux. The worldsheet theories are relatively conventional sigma model CFTs without the complication entailed by RR fluxes. The dual 2d spacetime CFTs are also much more tractable than their higher dimensional counterparts. The dictionary relates correlators of primary fields in the spacetime CFTs with those of physical vertex operators in the worldsheet CFT which are integrated then over the moduli space of Riemann surfaces with marked points,
\begin{multline} \label{corresp}
\int_{{\cal M}_{g,n}} \!\!\! \! \!\! \big\langle \mathcal{V}^{w_1}_{h_1}(x_1; z_1)\mathcal{V}^{w_2}_{h_2}(x_2; z_2) \ldots \mathcal{V}_{h_n}^{w_n}(x_n; z_n) \big\rangle_{\Sigma_{g,n}}  \\
= \big\langle {\cal O}^{(w_1)}_{h_1}(x_1){\cal O}^{(w_2)}_{h_2}(x_2) \ldots {\cal O}^{(w_n)}_{h_n}(x_n) \big\rangle_{\text{S}^2} \Big|_{g} \ .
\end{multline}
The spacetime CFT $n$-point function on ${\rm S}^2$ (the RHS) is organised in a genus expansion in a suitable large $N$ limit, and the genus-$g$ contribution is equated to the worldsheet CFT computation on a Riemann surface $\Sigma_{g,n}$ of genus-$g$ with $n$ marked points (the LHS).\footnote{The 
$\{h_i \}$ label conformal dimensions in the spacetime CFT, while the $\{ w_i\}$ are additional labels e.g.\ for the spectrally flowed sectors in the worldsheet CFT.} 
We might reasonably expect to be able to mathematically relate these two separately well-defined quantities. 

The spacetime CFTs on the RHS are believed to be deformations of a symmetric orbifold CFT. The simplest examples are for the backgrounds ${\rm AdS}_3\times {\rm S}^3\times \mathcal{M}_4$ with ${\cal M}_4=\mathbb{T}^4$ or ${\rm K3}$. Here the corresponding CFT is ${\rm Sym}^N({\cal M}_4)$ which involves taking $N$ copies of ${\cal M}_4$,
and orbifolding by the permutation group $S_N$ acting on these copies. To compare with perturbative string theory we take a large $N$ limit. The simplest case where we can address our question of mapping the two sides of eq.~\eqref{corresp} is at the actual orbifold point where the dual CFT is  tractable. 

Recently it has been proposed \cite{Gaberdiel:2018rqv,Eberhardt:2018ouy} that these orbifold CFTs are exactly dual to string backgrounds with pure NS-NS flux \cite{Maldacena:2000hw, Maldacena:2000kv, Maldacena:2001km}. The tensionless string case, which arises when the amount of NS-NS flux is minimal $k=1$ \cite{Gaberdiel:2017oqg,Ferreira:2017pgt}, is the simplest such instance where the dual orbifold CFT is simply ${\rm Sym}^N({\cal M}_4)$ with ${\cal M}_4=\mathbb{T}^4$ or ${\rm K3}$ \cite{Gaberdiel:2018rqv,Eberhardt:2018ouy}. Nontrivial evidence including a detailed match of the full perturbative string spectrum as well as the selection rules for three point correlators was provided \cite{Gaberdiel:2018rqv,Eberhardt:2018ouy}. Many of the considerations also generalise to the case with $k>1$ where it was seen that the perturbative long string spectrum matches precisely with that of the symmetric orbifold ${\rm Sym}^N(\widetilde{\mathcal{M}}_4)$ where $\widetilde{\mathcal{M}}_4$ has a Liouville factor in addition to  $\mathcal{M}_4$ \cite{Eberhardt:2019qcl}. It was also shown in 
 \cite{Eberhardt:2019qcl} that a DDF-like \cite{DelGiudice:1971yjh} construction on the worldsheet generates the correct spacetime symmetry algebra.  

In this paper we will identify a more structural reason underlying this agreement which goes quite some way towards establishing the mapping between the two sides of eq.~\eqref{corresp}. 
We will start with the LHS of eq.~(\ref{corresp}) for $g=0$ and provide strong evidence that, for the worldsheet path integral with $k=1$ units of NS-NS flux, the correlator of physical vertex operators delta-function-localises to a few discrete sets of points in the moduli space ${\cal M}_{0,n}$. These are the worldsheet Riemann surfaces which admit a holomorphic covering map to the spacetime ${\rm S}^2$ (the boundary of ${\rm AdS}_3$) such that the marked points $z_i$ on $\Sigma_{0,n}$ are  mapped to the $x_i$ on the spacetime ${\rm S}^2$ (with a specified branching $w_i$ at $z_i$). The worldsheet correlator, we claim, is thus generically zero apart from at these special points in moduli space!\footnote{These special points in moduli space are exactly where, for generic $k>1$, one found 
singularities in correlators (see Section~2.4 of \cite{Maldacena:2001km}) or in the one-loop partition function (see Section~4.1 of \cite{Maldacena:2000kv}).}

The primary evidence for this statement comes from a careful analysis of the Ward identities that are obeyed by the correlators of spectrally flowed primaries in the (bosonic) $\mathrm{SL}(2, \mathds{R})_{k+2}$  WZW model.  These Ward identities lead to a set of complicated recursion relations, which can be solved by correlators with this unusual delta-function-localisation property. This special solution exists provided that 
\be
\sum_{i=1}^n j_i=\frac{k}{2}(n-2)+1\ , \label{eq:j condition1}
\ee
where $j_i$ labels the Casimir of the  $\mathrm{SL}(2, \mathds{R})_{k+2}$  WZW primary, $\mathcal{C}(j)=-j(j-1)$. For $k=1$, the entire world-sheet spectrum comes from the bottom of the continuum with $j_i=\frac{1}{2}$ \cite{Eberhardt:2018ouy}, and thus (\ref{eq:j condition1}) is automatically satisfied for all $n$. Therefore all the physical vertex operators in the perturbative string spectrum at $k=1$ obey this condition. 

This result matches beautifully with some known facts about correlators in the orbifold CFT on the RHS. 
Lunin and Mathur, in a series of papers \cite{Lunin:2000yv,Lunin:2001pw}, showed how one can compute correlators (of twisted sector states) in a symmetric orbifold CFT in terms of an auxiliary covering surface with a branching structure over the points $x_i$ dictated by the (single-cycle) twists $w_i$ of the operators. They noticed that this auxiliary surface had a nontrivial genus and that it seemed to play the role of the dual worldsheet. This conjecture was formalised in works of Pakman, Rastelli and Razamat \cite{Pakman:2009zz,Pakman:2009ab} where they elucidated this covering space picture from the point of view of the orbifold CFT. At the time there was no comparison to a bonafide worldsheet theory. Our results show that this picture is indeed realised at the level of the worldsheet theory for NS-NS backgrounds, in the tensionless limit of $k=1$. The nontrivial localisation of the worldsheet path integral, that this picture demands, is something we now see explicitly from the sigma model, at least for this special solution. 

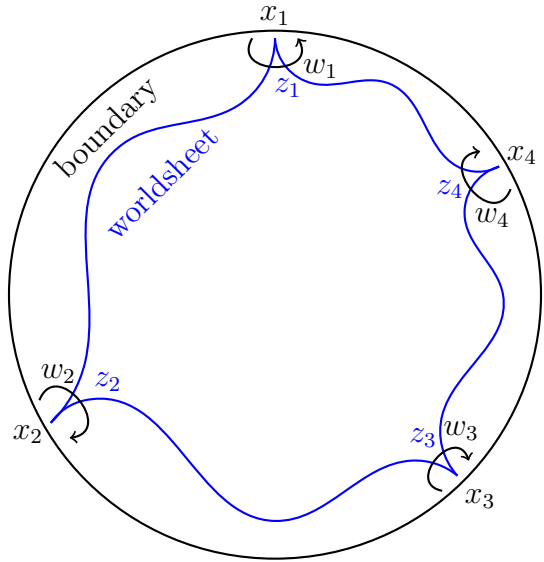
\begin{figure}
\begin{center}
\begin{tikzpicture}
\draw[thick] (0,0) circle (3.5);
\draw[thick,blue] (0,3.4) to [curve through={(-.5,2.5) .. (-2,2) .. (-2.5,-.9)}] (-2.95,-1.7);
\draw[thick, blue] (-2.95,-1.7) to [curve through={(-2.5,-1.4) .. (0,-3) .. (1.8,-2.2)}] (2.4,-2.4);
\draw[thick,blue]  (2.4,-2.4) to [curve through={(2.2,-1.6) .. (3,0) .. (2.5,1)}] (2.95,1.7);
\draw[thick,blue]  (2.95,1.7) to [curve through={(2.4,1.7) .. (1.5,2.8) .. (.5,2.8)}] (0,3.4);
\draw[thick,->,bend left=120,looseness=2.5] (2.2,-2.6) to (2.55,-2.2);
\node[rotate=45,blue] at (-1.5,1.5) {worldsheet};
\node[rotate=45] at (-2.2,2.2) {boundary};
\draw[thick,->,bend right=120,looseness=2.5] (-.3,3.4) to (.3,3.4);
\draw[thick,->,bend left=120,looseness=2.5] (-3.1,-1.4) to (-2.7,-1.9);
\draw[thick,->,bend left=120,looseness=2.5] (3.1,1.4) to (2.7,1.9);
\node at (0,3.7) {$x_1$};
\node at (-3.25,-1.85) {$x_2$};
\node at (2.7,-2.7) {$x_3$};
\node at (3.25,1.85) {$x_4$};
\node[blue] at (.17,2.75) {$z_1$};
\node[blue] at (-2.2,-1.15) {$z_2$};
\node[blue] at (1.95,-1.9) {$z_3$};
\node[blue] at (2.3,1.4) {$z_4$};
\node at (.6,3) {$w_1$};
\node at (2.45,-1.8) {$w_3$};
\node at (-2.85,-1.05) {$w_2$};
\node at (2.86,1.02) {$w_4$};
\end{tikzpicture}
\end{center}
\caption{A cartoon of the semiclassical string solution in the example of a 4-point function. The worldsheet is pinned to the boundary sphere via the covering map and has ramification indices $w_i$ at the insertion points.}\label{fig:4pt function}
\end{figure}

While the Ward identities do not uniquely pick out this particular solution, additional support that the delta-function-localised worldsheet correlator is indeed the correct one comes from a semiclassical picture. We argue  that classical string solutions corresponding to states at the bottom of the continuum, i.e.\ with $j_i=\frac{1}{2}$, are essentially pinned to the $\text{S}^2$ boundary of the ${\rm AdS}_3$. (As we mentioned before, these correspond to all the physical states that are present for $k=1$ \cite{Eberhardt:2018ouy}.) The solutions corresponding to spectrally flowed states in a sector labelled by an integer $w$ have locally a behaviour which is that of a $w$-branched cover of the boundary $\text{S}^2$. We also find an exact classical solution corresponding to the global holomorphic covering map from the worldsheet to $\text{S}^2$ that contributes to an $n$-point correlator. A schematic picture of these classical solutions is given in Figure~\ref{fig:4pt function}. The action of this semi-classical configuration agrees exactly with the Liouville action that appears in the Lunin-Mathur computation of the orbifold CFT correlator. Moreover, since these solutions are localised at the boundary of ${\rm AdS}_3$, one can argue that the semiclassical approximation should be exact! 

The overall picture that emerges is that correlators of physical vertex operators in the $k=1$ theory 
have the rather remarkable feature that they are localised onto a set of semiclassical configurations corresponding to holomorphic covering maps of the boundary of ${\rm AdS}_3$  by the worldsheet. 
We see this from the Ward identities of the sigma model CFT, as well as from a semiclassically exact analysis of classical string solutions. This worldsheet CFT analysis reproduces all the features of the Lunin-Mathur approach to the computation of the spacetime CFT correlators. Together with the results in 
\cite{Gaberdiel:2018rqv,Eberhardt:2018ouy}, we believe this essentially demonstrates how the string theory on  ${\rm AdS}_3\times {\rm S}^3\times {\cal M}_4$ at $k=1$ becomes equivalent to the spacetime orbifold CFT ${\rm Sym}^N({\cal M}_4)$ in a large $N$ expansion. 
    
\smallskip

The organisation of this paper is as follows. In Section~\ref{sec:overview} we set out in more detail the general structure of our argument and give markers for how the different pieces of the analysis fit together. Sections~\ref{sec:setup}--\ref{sec:simple solution} discuss the world-sheet approach to calculating the correlation functions. More specifically, after defining our conventions and explaining a few basic facts about spectrally flowed vertex operators in Section~\ref{sec:setup}, we explain how to formulate the Ward identities for their correlators in Section~\ref{sec:Ward}. Section~\ref{sec:simple solution} describes then the, in general, delta-function-localised solutions of these Ward identities.

In preparation for the comparison to the classical solutions, we explain in Section~\ref{sec:covering} how these correlation functions can be expressed in terms of the Wakimoto free field realisation for which the relation to the covering map becomes manifest. In Section~\ref{sec:classical} we discuss the relevant classical solutions and show that they match precisely with the results from Section~\ref{sec:covering}, thus linking our quantum analysis to the Lunin-Mathur description of symmetric orbifold correlators. Finally, we sketch in Section~\ref{sec:genus} how one should expect the analysis to generalise to arbitrary genus on the world-sheet, and we summarise our findings and suggest further directions for the future in Section~\ref{sec:concl}.

There are three appendices: Appendix~\ref{app:npointsol}  gives the detailed argument that our ansatz for the correlators in the general case (see Section~\ref{sec:gencorr}) satisfies indeed the Ward identities of Section~\ref{sec:Ward}, while Appendices~\ref{app:secret} and \ref{app:anomalous partialPhi} explain various  technical issues that arise in the context of the Wakimoto free field realisation of Section~\ref{sec:covering}.

\section{From Worldsheet CFT to Spacetime CFT}\label{sec:overview}

This section firstly gives an overview of the big goal as well as the concrete 
strategy we adopt towards realising it. It then reviews some of the ingredients in this strategy, in particular, the Lunin-Mathur approach to computing correlators in symmetric orbifold CFTs. 

\subsection{Goal and Roadmap}\label{sec:Goal}

As mentioned in the introduction, the big goal is to derive the ${\rm AdS}_3/{\rm CFT}_2$ correspondence (at $k=1$) by recasting the correlators of the worldsheet CFT (the LHS of eq.~(\ref{corresp})) in a way which makes them manifestly equivalent to the spacetime CFT correlators (the RHS of eq.~\eqref{corresp}). Showing the equality in this way would then bypass explicit computation of individual correlators and their comparison. The latter, is in any case, usually prohibitive beyond the simplest cases and not always particularly illuminating. We will focus largely on the $k=1$ tensionless limit where the orbifold CFT is relatively simple and we do not have the complications arising from the presence of the long string continuum \cite{Seiberg:1999xz, Maldacena:2000hw}.\footnote{Much of the quantum analysis will be carried out for general $k$, and some of these results are likely to be useful in studying the general case. In fact, these give a systematic way to study correlators of spectrally flowed operators through the Ward identity constraints. However, as we will see in Section~\ref{sec:jcond} and later, there are major simplifications when the condition of eq.~(\ref{eq:j condition1}) holds, and we will restrict to that case which is in anyway of direct relevance to the $k=1$ theory.}

The strategy we will adopt is twofold. The bulk of the paper (Sections~\ref{sec:setup}--\ref{sec:simple solution}) is devoted to a CFT analysis of the  
correlators of spectrally flowed states in the bosonic $\mathrm{SL}(2, \mathds{R})_{k+2}$  WZW model. These are the building blocks in any covariant analysis of the worldsheet theory on 
${\rm AdS}_3$ as well as for the evaluation of the string amplitudes defined on the LHS of eq.~(\ref{corresp}). 
This is the case, irrespective of whether one uses the NSR formulation of \cite{Maldacena:2000hw,Maldacena:2000kv,Maldacena:2001km} (or rather its supersymmetric extension developed in \cite{Giveon:1998ns,Israel:2003ry,Raju:2007uj,Ferreira:2017pgt}), or the hybrid formalism of \cite{Berkovits:1999im} that was employed in \cite{Eberhardt:2018ouy}. (In either case, one can decouple the fermions and deal with the decoupled bosonic $\mathfrak{sl}(2, \mathds{R})_{k+2}$ algebra at level $k+2$.)

For simplicity, we will concentrate on vertex operators for the physical ground state in each spectrally flowed sector. From the detailed matching of the spectrum between the symmetric product CFT and the string theory, we know that the $w$-spectrally flowed sector corresponds to the $w$-cycle twisted sector in the orbifold \cite{Gaberdiel:2018rqv,Eberhardt:2018ouy}, see also \cite{Giveon:2005mi,Giribet:2018ada}. We thus have a correspondence
\be
\mathcal{V}^{w}_{h}(x; z) \longleftrightarrow {\cal O}^{(w)}_{h}(x) \ .
\ee
Here the spacetime CFT operator on the RHS is the twist field creating the ground state in the $w$-twisted sector. The operator on the LHS, on the other hand, creates the state $|j=\frac{1}{2}, m \rangle^{(w)}$ which sits at the bottom of the continuous representations in the $w$-spectrally flowed sector of the $\mathrm{SL}(2, \mathds{R})_{k+2}$ WZW model. The $J^3_0$ quantum number $m$ is fixed by the physical state conditions and related to the spacetime conformal dimension as $h=m+\frac{k+2}{2}w$. 
Since we are considering the ground states, we will not really need the rest of the contributions (i.e.\ the degrees of freedom from $\mathrm{S}^3\times \mathcal{M}_4$ as well as the ghosts) to the full string theory vertex operator. We will thus denote the $\mathfrak{sl}(2, \mathds{R})$ vertex operators, which we will focus on, by  
$V_{h}^{w}(x;z)$ to distinguish them from the full string vertex operators ${\cal V}^{w}_{h}(x; z)$ that appear in  eq.~(\ref{corresp}). 

The main aim of the world-sheet analysis is to calculate the (unintegrated) correlators on the sphere
\be\label{gencorr} 
\langle V_{h_1}^{w_1}(x_1;z_1)V_{h_2}^{w_2}(x_2;z_2) \cdots V_{h_n}^{w_n}(x_n;z_n) \rangle \ ,
\ee
and constrain them using the Ward identites obtained by additional $\mathfrak{sl}(2, \mathds{R})_{k+2}$ current insertions $J^a(z)$.\footnote{As in \cite{Maldacena:2001km}, we will be considering the Euclidean Wick rotation on both the spacetime as well as the worldsheet. Thus strictly speaking, we are looking at $\mathfrak{sl}(2, \mathds{C})$ Ward identities. The distinction is immaterial since we are looking only at the chiral sector.} Unlike in the case of the spectrally unflowed sector, where this reduces to the action of zero modes, here near each $z_i$ the first $w_i$ positive modes of $J^+$ act nontrivially since they do not annihilate the spectrally flowed ground state, see in particular eq.~(\ref{eq:JpVn}) below. This is compensated by the fact that for a suitable combination involving $J^-$ the situation is reversed, and one has a zero of order $w_i-2$ near $z_i$, see eq.~(\ref{eq:J-const}). This allows us to evaluate the current correlators in terms of a number of unknown coefficients (see eq.~(\ref{eq:hatF})), which however satisfy corresponding constraints.  

Unlike in the unflowed sector, solving this set of unknowns and constraints is non-trivial in general. Nevertheless, we find a rather remarkable solution to this set of equations which are determined in terms of the data of a holomorphic covering map
\be\label{covmap}
x= \Gamma(z) \ ,
\ee
which maps the worldsheet to the target $\mathrm{S}^2$ and has branching number $w_i$ at $z=z_i$. (We review basic facts about covering maps in Section~\ref{sec:gencov}.) However, once one fixes three points in the target space, the covering map specifies the other points $x_i$ in the image of the $(n-3)$ worldsheet points $z_i$. (Alternatively, the covering map specifies the remaining world-sheet insertion points $z_i$ in terms of the spacetime insertion points $x_i$). In fact, we find that the correlator in (\ref{gencorr}) has $(n-3)$ delta functions $\delta(x_i-\Gamma(z_i))$, see eq.~(\ref{eq:localisation solution}). This is a rather unusual behaviour for a correlator in 
a 2-dimensional CFT.\footnote{Note that the $\mathfrak{sl}(2,\mathds{R})_{k+2}$ WZW model is non-unitary, so this is not in conflict with constraints like reflection positivity. We should also mention that a similar delta function localisation was already observed in the $k=1$ one-loop worldsheet partition function in \cite{Eberhardt:2018ouy}.} Therefore it is not surprising that this solution does not generically exist --- we need the condition (\ref{eq:j condition1}) to be satisfied.  As mentioned in the introduction, this condition is satisfied for the physical states at the bottom of the continuum i.e.\ for $j_i=\frac{1}{2}$ at $k=1$ (and any value of $h_i$). 
\medskip

This is a particular solution of the Ward identities. That it is the relevant one, is buttressed by the second prong of our analysis which starts with the semiclassical description of  string theory on $\mathrm{AdS}_3$ for which the bosonic degrees of freedom are controlled by the effective action 
\be\label{adsact}
S_{\mathrm{AdS}_3}= \frac{k}{4\pi} \int \mathrm{d}^2 z\, \sqrt{g}\,  \Bigl( 4\,  \partial \Phi \, \bar\partial \Phi +\beta\,\bar{\partial}\gamma+\bar{\beta}\,\partial \bar{\gamma}- \mathrm{e}^{-2\Phi} \beta\bar{\beta}  \Bigr)\ .
\ee
In this parametrisation of $\mathrm{AdS}_3$, $\Phi$ is the radial coordinate and $\gamma$ and $\bar{\gamma}$ are the coordinates of the boundary sphere \cite{Giveon:1998ns}. The boundary of $\mathrm{AdS}_3$ is located at $\Phi \to \infty$. 

We will start with the semiclassical solutions which correspond to the ground states of the spectrally flowed sector i.e. sit at the bottom of the continuum. These arise in taking a certain scaling limit of a family of solutions. The end result (the details will be described in Section~\ref{subsec:classical sol} below) is a solution
\begin{subequations} \label{coversol}
\begin{align}
\Phi(z,\bar{z}) & =  -\log(\epsilon) -  \tfrac{(w-1)}{2}\log(\bar{z}) - \tfrac{(w-1)}{2}\log(z) \ , \\
\gamma (z) & = z^w \ .
\end{align}
\end{subequations}
Here $\epsilon$ is a parameter that goes to zero as one takes the scaling limit. What this means is that the radial coordinate $\Phi$ of the string worldsheet is essentially at the boundary ($\Phi\rightarrow \infty$) in this limit. Furthermore, the local map from the worldsheet to the target $\mathrm{S}^2$ parametrised by $\gamma$ is holomorphic, and has a branching of order $w$ (near $z=0$). What is important here is that for wavefunctions localised at the boundary, the semiclassical description is exact (see for example \cite{deBoer:1998gyt}) even though we are considering a highly curved spacetime at $k=1$. Furthermore, the worldsheet theory actually becomes free in this limit. 
This is essentially because in writing the action (\ref{adsact}) in first order form, see eq.~(\ref{adsact2}),  the nontrivial interaction term vanishes when $\Phi\rightarrow \infty$. Put differently, only when $\gamma$ is holomorphic, $\Phi$ can reach the boundary and the action \eqref{adsact} stays finite.

In fact, as we will see in Section~\ref{sec:classical}, 
one can find a classical solution which describes the general correlator in (\ref{gencorr}) and has the right behaviour as in (\ref{coversol}) near each of the $z_i$. It is given in terms of the global covering map (\ref{covmap}) and therefore exists only if $x_i=\Gamma(z_i)$ --- in particular, the correlator (with the correct behaviour near each of the insertions) can only be nonzero if this covering map exists. Furthermore, based on the Wakimoto realisation of Section~\ref{sec:covering}, this classical solution agrees precisely with the special solution of the WZW model found in Section~\ref{sec:simple solution} (where the solutions were delta-function-localised to the points in moduli space where the covering map exists). This justifies {\it post hoc}, that the solution of the Ward identities of Section~\ref{sec:simple solution} is indeed the right one.\footnote{A similar argument for $k>1$ shows that for states at or near the bottom of the continuum, the classical solutions are localised near the boundary and one can use semiclassical arguments. However, now the presence of the continuum (which was absent for $k=1$) leads to an additional integral over the radial momentum, and this smears the delta-functions into pole-like singularities as in the discussion around eq.~(2.35) of \cite{Maldacena:2001km}. Note that their reasoning should also hold in the spectrally flowed sector, as long as the states are near the bottom of the continuum.}

This outlines the strategy to show that the unintegrated vertex operators on the LHS of eq.~(\ref{corresp}) are 
delta-function-localised to the isolated points which admit holomorphic covering maps to the target space. 
Once we integrate over the moduli space we then pick up these localised contributions. 
Because the theory is semiclassically exact, the correlator itself is then obtained by evaluating the action on-shell with these (isolated) covering map solutions. 
We find that this semiclassical contribution matches exactly with the Lunin-Mathur approach to computing the twisted sector ground state correlators in symmetric orbifold CFTs, as we shall now review. 

\subsection{Review of the Lunin-Mathur Construction}

Correlators in orbifold theories can be evaluated by going to the covering space \cite{Hamidi:1986vh, Dixon:1986qv}. In the special case of 
symmetric product orbifolds of a seed CFT, Lunin and Mathur \cite{Lunin:2000yv}, have an elegant way of implementing this in a path integral approach. 

\begin{figure}
\begin{center}
\begin{minipage}{.49\textwidth}
\includegraphics[width=\textwidth]{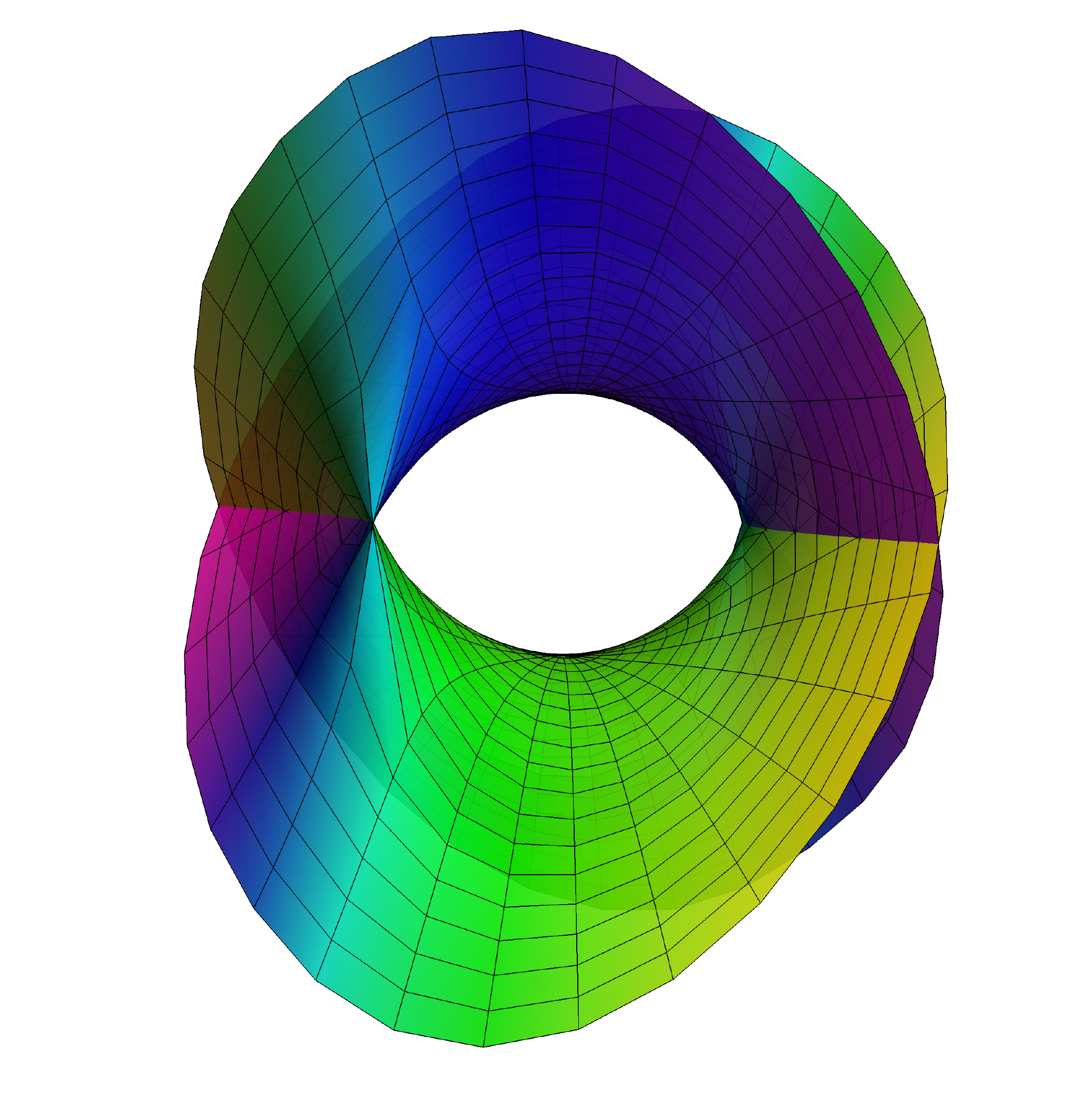}
\end{minipage}
\begin{minipage}{.49\textwidth}
\includegraphics[width=\textwidth]{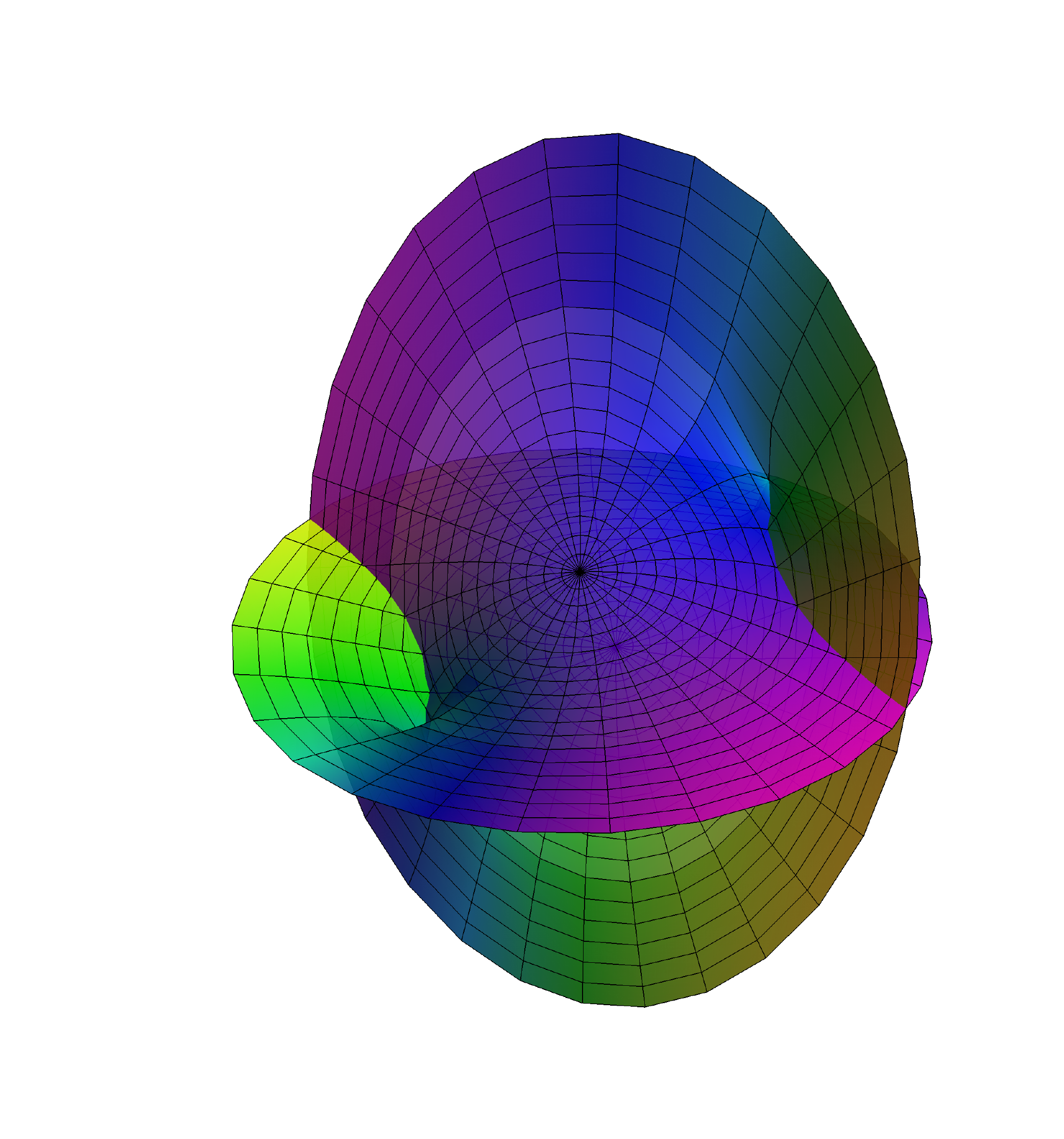}
\end{minipage}
\end{center}
\caption{A covering surface with two branch points of order $2$. We plot the real part of the inverse of the covering map, which has two sheets. We show the two sheets from a side view (left) and a top view (right). There are two branch points in the picture and monodromy around them interchanges the two sheets.} \label{fig:covering map}
\end{figure}

Very briefly, the idea is that, restricting to correlators of twist fields without excitations, we only need to evaluate a path integral on the covering space with specified branching at the corresponding branch points. The effect of the twist field, which is to introduce a multi-valuedness on the fields of the seed CFT, is taken into account by defining a single valued field in the covering space. Thus a twist field corresponding to a $w$-cycle twist in $S_N$  is associated with a branch point of order $w$ of an $N$-sheeted covering of the original spacetime (an $\text{S}^2$) that the CFT lived on. A correlator of twist fields as in the RHS of (\ref{corresp}) is then evaluated by going to a covering space determined by the covering map which has the right branching behaviour $w_i$ at each pre-image of the point $x_i$ on the spacetime $\text{S}^2$. There are no insertions now in the covering space path integral and only one copy of the seed CFT.  We plotted one example of such a covering map in Figure~\ref{fig:covering map} for the case of two twist-2 fields, leading to two branch points of order $2$.

To define the path integral on the covering space requires a careful analysis  of cutting out open discs around the insertion points as well as at infinity, and imposing appropriate boundary conditions \cite{Lunin:2000yv}. These act as regulators in the path integral, and it can be shown that the final answer is independent of the exact choices made here. We will refer the reader to \cite{Lunin:2000yv} for the details of this, and will implicitly assume this as being done whenever we talk of evaluating the path integrals. 
\smallskip

To be more specific, let us denote the 
coordinates on the original CFT spacetime by $x$, and those on the  covering space (soon to be identified with the worldsheet) by $z$. For the case of ${\rm Sym}^N({\cal M}_4)$, we then have a path integral of the seed CFT ${\cal M}_4$, but now lifted to the covering space. There is a nontrivial metric on the covering space, which is induced from the branched cover $x=\Gamma(z)$. This is given by a conformal factor 
\be\label{covscale} 
\mathrm{e}^{\phi}=\big| \partial_z \Gamma\big|^2 \ .
\ee
Thus the partition function of $\mathcal{M}_4$, evaluated on the covering space, gets an additional Liouville term from this scale factor $\phi$, see \cite[eq.~(13.2)]{Friedan:1982is}
\be\label{liouv}
S_\text{L}[\phi] = \frac{c}{48\pi} \int \mathrm{d}^2 z\,  \sqrt{g}\, \Bigl(2 \, \partial \phi \, \bar\partial \phi+R\, \phi\Bigr) \ . 
\ee
Note that near a branch point of order $w$, we locally have $\Gamma(z) \sim z^w$ and thus $e^{\phi} \rightarrow 0$ (for $w>1$). All the nontrivial spacetime dependence of the correlators comes from this ``on-shell action" and is encoded in the covering map. In general, the covering map is not very easy to write down explicitly. However, we will not really need the actual form of the covering map to show that the correlators of the spacetime CFT, evaluated using the Lunin-Mathur approach, agree with the worldsheet. 

This is because, for the classical solution given in Section~\ref{sec:classical},
the $\text{AdS}_3$ radial field $\Phi(z)$ is essentially the same as the conformal factor $\phi(z)$ in (\ref{covscale})
\be\label{scalematch}
\partial\phi(z)=-2\partial\Phi(z)\ ,
\ee
see the discussion in Section~\ref{sec:onshell}. 
Furthermore, as already mentioned above, the field $\gamma(z)$ in (\ref{adsact}) is holomorphic (and in fact equal to $\Gamma(z)$), and hence does not contribute to the on-shell action of the worldsheet. With this identification, the Liouville action  (\ref{liouv}) then agrees exactly with the on-shell ${\rm AdS}_3$ action (\ref{adsact}) as rewritten in first order form, see eq.~(\ref{adsact2}), 
for $k=1$ (with $c=6k$).\footnote{The linear dilaton term in the $\text{AdS}_3$ action is generated through renormalisation at the quantum level \cite{Giveon:1998ns}.}

Thus the worldsheet computation of the correlator reduces, because of semi-classical exactness, to that of the on-shell action on this solution, which in turn agrees with that of the Liouville action in (\ref{liouv}) provided we identify the worldsheet with the covering space introduced by Lunin and Mathur. In fact, this was conjectured by Lunin-Mathur  and later elaborated on in 
\cite{Pakman:2009zz}. We now see a full realisation of this idea in our present setup where we have started with a bonafide worldsheet description of the $k=1$ string theory on $\text{AdS}_3$. 

This concludes our sketch of how we can manifestly exhibit the equivalence between the two sides of (\ref{corresp}), at least for genus zero. We will also briefly describe in Section~\ref{sec:genus}, why we expect the considerations to go through even for higher genus.

\section{The basic worldsheet CFT setup}\label{sec:setup}

In this section we begin with the worldsheet analysis of the correlators in eq.~(\ref{corresp}). Thus we will study the correlation functions of spectrally flowed affine primaries of the bosonic $\mathrm{SL}(2,\mathds{R})_{k+2}$ WZW model. The reason why we are interested in these states is that they correspond to the ground states of the corresponding twisted sector of the dual  symmetric orbifold theory \cite{Gaberdiel:2018rqv,Eberhardt:2018ouy}. 

For the following it will be very important to understand precisely how the relevant vertex operators (that we shall denote as $V^w_h(x;z)$, where $w$ labels the spectral flow sector) are defined. From the viewpoint of the dual CFT the ground states of the twisted sectors are quasiprimary --- in fact, they are even primary --- with conformal weight $h$, and thus the associated vertex operators transform under the action of the global M\"obius generators of the spacetime CFT (whose coordinates we shall denote by $x$) as 
\be
[ J^a_0, V^w_h(x;z) ] = -\mathcal{D}^a \, V^w_h(x;z) \ , 
\ee
where the $\mathcal{D}^a$ are the differential operators 
\be \label{eq:xbasis}
\mathcal{D}^+ = -\frac{\partial}{\partial x}\ , \qquad \mathcal{D}^3=-h -x \frac{\partial}{\partial x}\ , \qquad \mathcal{D}^-=-2 h x -x^2 \frac{\partial}{\partial x}\ . 
\ee
This requirement fixes the $x$-dependence of the vertex operators; in particular, since $J^+_0$ is the translation operator in $x$-space, we have 
\be \label{eq:Vtrans}
V_h^w(x+y;z+\zeta)=\mathrm{e}^{y J_0^+} \mathrm{e}^{\zeta L_{-1}} \, V_h^w(x;z)\, \mathrm{e}^{-\zeta L_{-1}}\mathrm{e}^{-y J_0^+} \ ,
\ee
where we have also used that the translation operator on the worldsheet, i.e.\ in $z$, is $L_{-1}$. Note that $L_{-1}$ and $J^+_0$ commute, and thus there is no ordering ambiguity. 

Since the $\mathfrak{sl}(2,\mathds{R})$ zero modes change the value of $x$, we may think of $x$ (from a worldsheet perspective) as describing the different states in a given  $\mathfrak{sl}(2,\mathds{R})$ representation; this viewpoint is sometimes referred to as the `$x$-basis'. We should also mention that the Casimir of this representation is 
\be
C = - (\mathcal{D}^3)^2 + \tfrac{1}{2} \bigl( \mathcal{D}^+ \mathcal{D}^- + \mathcal{D}^- \mathcal{D}^+ \bigr) = - h (h-1) \ . 
\ee
\smallskip

For the following it will be important to understand the structure of the OPE of the affine $\mathfrak{sl}(2,\mathds{R})_{k+2}$ currents with these vertex operators. We identify fields and states at $(x;z)=(0;0)$, and on these states the spectral flow automorphism of the affine $\mathfrak{sl}(2,\mathds{R})_{k+2}$ algebra is defined to be, see e.g.\ \cite{Eberhardt:2019qcl}
\begin{align}
\sigma^w(J^\pm)(z)&=z^{\mp w} J^\pm(z)\ , \\
\sigma^w(J^3)(z)&=J^3(z)+\frac{(k+2) w}{2z}\ .
\end{align}
For an affine primary field in the $w$'th spectrally flowed sector the OPEs are thus
\begin{subequations}\label{eq:def OPEs V}
\begin{align} 
J^+(z) V_h^w(0;0)&\sim \sum_{p=2}^{w+1} \frac{(J^+_{p-1} V_h^w)(0;0)}{z^p}+\frac{\partial_x V_h^w(0;0)}{z}\ , \label{eq:def OPEs V a}\\
J^3(z) V_h^w(0;0) &\sim \frac{h V_h^w(0;0)}{z}\ , \label{eq:def OPEs V b}\\
J^-(z) V_h^w(0;0) &\sim \mathcal{O}(z^{w-1})\ , \label{eq:def OPEs V c}
\end{align}
\end{subequations}
where in \eqref{eq:def OPEs V a} we have separated out  the term involving $J^+_0$, which can be identified with the differential operator because of \eqref{eq:xbasis}. We should mention that the last OPE \eqref{eq:def OPEs V c} is not only regular, but that the first $w-1$ regular terms vanish. On the other hand, in \eqref{eq:def OPEs V a} the positive modes $J^+_{p-1}$ generically do not annihilate the spectrally flowed state, since they act as $\sigma^w(J^+_{p-1}) = J^+_{p-1-w}$, and hence we typically have a pole of order $(w+1)$.  

\begin{figure}
\begin{center}
\begin{tikzpicture}
\draw[thick,->] (4,.5) -- (5,.5) node[right] {$J^3_0$};
\draw[thick,->] (4,.5) -- (4,1.5) node[above] {$L_0$};
\node (A) at (-2,2) {};
\node (B) at (-1,1) {};
\node (B1) at (-1,1.5) {};
\node (C) at (0,0) {};
\node (C3) at (0,1.5) {};
\node (D) at (1,-1) {};
\node (D5) at (1,1.5) {};
\draw[red, fill=red!50!white, opacity=.2, rotate around={-45:(-2,2)}] (-2.8,1.9) rectangle (2.6,2.1);
\draw[line width=3, white] (-2.8,2.3) -- (-2.3,2.8);
\draw[line width=3, white] (1,-1.5) -- (2,-.5);
\foreach \i in {-2,...,1}{
\pgfmathtruncatemacro\ii{2*\i}
\foreach \j in {-\ii,...,5}{
\pgfmathparse{.5*\j}
\let\halfj\pgfmathresult
\node[circle, fill,draw,scale=.3] at ({\i},{\halfj}) {};
}
}	
\draw[bend right=30,->] (C) to node[left] {$J^+_2$} (D);
\draw[bend left=30,->] (C) to node[left] {$J^-_{-2}$} (B);
\draw[bend right=30,->] (A) to node[above,xshift=1ex] {$J^+_1$} (B1);
\draw[bend right=30,->] (B1) to node[below] {$J^+_0$} (C3);
\draw[bend right=30,->] (C3) to node[below] {$J^+_0$} (D5);
\end{tikzpicture}
\end{center}
\caption{The spectrally flowed representation of $\mathfrak{sl}(2,\mathds{R})_{k+2}$ in the example $w=2$. The `edge states' (inside the red strip) are the spectrally flowed affine primaries and they form an $\mathfrak{sl}(2,\mathds{R})$ representation under $J^\pm_{\pm w}$ and $J^3_0-\frac{(k+2)w}{2}$, see eq.~\eqref{zeromode}. The states away from the edge have higher multiplicity and can be reached with the affine oscillators. The global $\mathfrak{sl}(2,\mathds{R})$ zero-mode algebra acts horizontally and is resummed in the $x$-basis.}\label{fig:module}
\end{figure}
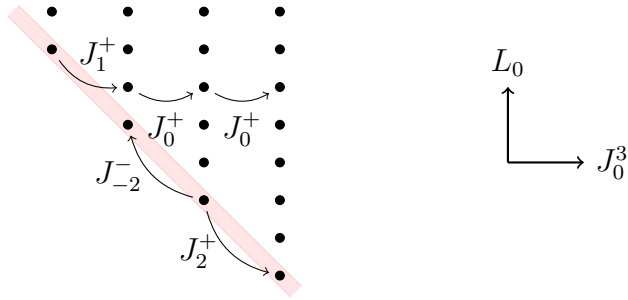

We should note that the vertex operators $V_h^w(x;z)$ also depend on the quantum number $j$ that we often suppress in our notation. To understand the origin of this quantum number we recall that the modes 
\be\label{zeromode}
\tilde{J}^+_0 = J^+_w \ , \qquad \tilde{J}^{3}_0 = J^3_0-\tfrac{(k+2)w}{2} \ , \qquad \tilde{J}^-_0 = J^-_{-w} 
\ee
form an $\mathfrak{sl}(2,\mathds{R})$ algebra, which is to be identified with the zero mode algebra before spectral flow. These zero modes act on a highest weight representation (before spectral flow), and the spin of this representation will be denoted by $j$. We will always work with the conventions that 
\begin{subequations}\label{eq:sl2action}
\begin{align}
\tilde{J}^+_0 |j,m\rangle & = (m+j) \, |j,m+1\rangle \label{eq:J+zero} \\
\tilde{J}^3_0 |j,m\rangle & = m \, |j,m\rangle \label{eq:J3zero} \\
\tilde{J}^-_0 |j,m\rangle & = (m-j) \, |j,m-1\rangle \ . \label{eq:J-zero}  
\end{align}
\end{subequations}
We do not put any restrictions on $j$ and assume generically that the representation does not truncate (as can happen for $j \in \mathds{R}$). 
Note that it follows from the middle equation of (\ref{zeromode}) together with (\ref{eq:J3zero}) that the parameter $m$ here is related to the conformal weight by, see also \cite[eq.~(2.14)]{Gaberdiel:2018rqv}
\be 
h=m+\tfrac{(k+2)w}{2}\ . \label{eq:hm relation}
\ee 
Finally, the worldsheet conformal dimension takes the form 
\be\label{Deltadef}
\Delta = - \frac{j(j-1)}{k}-w h  + \frac{(k+2)w^2}{4}  \ ,
\ee
but it will not play an important role in this paper.
A schematic picture of these modules is given in Figure~\ref{fig:module}.
\smallskip

The above OPE relations of eq.~(\ref{eq:def OPEs V}) specify the OPEs for states inserted at $(x;z)=(0;0)$, and the general case can now be obtained from this using \eqref{eq:Vtrans}. In particular, we note that 
\begin{align}\label{eq:transprop}
J^a(\zeta) \, V^{w}_h(x;z) & = J^a(\zeta) \, \mathrm{e}^{x J_0^+} \, V_h^w(0;z)\, \mathrm{e}^{- x J_0^+} \nonumber \\
& = \mathrm{e}^{x J_0^+} \bigl[ J^{a (x)}(\zeta) \, V_h^w(0;z)\bigr] \mathrm{e}^{- x J_0^+} \ , 
\end{align}
with
\begin{subequations}\label{eq:Jxshift}
\begin{align}
J^{+(x)}(z) & = J^+(z) \ , \label{eq:J+x}\\
J^{3(x)}(z) & = J^{3}(z) + x J^+(z) \ ,\label{eq:J3x}  \\ 
J^{-(x)}(z) & = J^-(z) + 2 x J^3(z) + x^2 J^+(z) \ . \label{eq:J-x}
\end{align}
\end{subequations}
We shall use these OPEs in the following chapter to derive Ward identities for the corresponding correlation functions.

\section{The Ward identities}\label{sec:Ward}

In this section we want to determine the Ward identities for the correlation functions 
\be 
\langle V_{h_1}^{w_1}(x_1;z_1)V_{h_2}^{w_2}(x_2;z_2) \cdots V_{h_n}^{w_n}(x_n;z_n) \rangle
\ee
on the sphere. Here $V_{h}^{w}(x,z)$ is the $w$-spectrally flowed image of an affine primary vertex operator as described above. If all $w_i$ are equal to zero, the Ward identities are well-known, 
\be 
\left\langle J^a(z)\prod_{i=1}^n V_{h_i}^{0}(x_i;z_i) \right\rangle=-\sum_{i=1}^{n} \frac{\mathcal{D}^a_i}{z-z_i} \left\langle \prod_{j=1}^n V_{h_j}^{0}(x_j;z_j) \right\rangle\ . \label{eq:unflowed Ward identity}
\ee
Here we have used that all $V_{h_i}^{0}(x_i;z_i)$ are highest weight operators, and hence the OPE of $J^a(z)$ with any such operator has only a first order pole, which is given by the right-hand-side of \eqref{eq:unflowed Ward identity}. As we have seen above, this is no longer true for $w>0$ since then $J^+(z)$ has a pole of order $w+1$, and the same is also true for the other $J^a(z)$ because of \eqref{eq:Jxshift}. 
In the following we want to find the spectrally flowed analogue of these identities.

\subsection{The constraint equations}\label{sec:basic strategy}

The direct analogue of \eqref{eq:unflowed Ward identity} for $a=+$ is 
\begin{multline} 
\left\langle J^+(z) \prod_{i=1}^n V_{h_i}^{w_i}(x_i;z_i) \right \rangle=\sum_{i=1}^n \frac{\partial_{x_i}}{(z-z_i)} \left\langle  \prod_{l=1}^n V_{h_l}^{w_l}(x_l;z_l) \right \rangle \\
+ \sum_{i=1}^{n} \sum_{\ell=1}^{w_i} \frac{1}{(z-z_i)^{\ell+1}} \left\langle (J^+_\ell V_{h_i}^{w_i})(x_i;z_i) \prod_{l\ne i} V_{h_l}^{w_l}(x_l;z_l) \right \rangle \ , \label{eq:JpVn}
\end{multline}
because the right-hand-side accounts for all the poles (in $z$) of the left-hand-side, see eq.~(\ref{eq:def OPEs V a}). However, as it stands, this is not particularly useful because we do not know how to evaluate the terms 
\be\label{eq:hatF}
\hat{F}^i_\ell = \Bigl\langle (J^+_\ell V^{w_i}_{h_i}) (x_i;z_i) \, \prod_{l\neq i} V_{h_l}^{w_l}(x_l;z_l) \Bigr\rangle 
\ee
for $\ell=1,\ldots,w_i$ that appear in the second line. In order to determine them, we take a small detour and first compute the correlator with $J^3(z)$ or $J^-(z)$ inserted instead of $J^+(z)$. Using \eqref{eq:Jxshift} and accounting again for the poles in $z$ (using eqs.~\eqref{eq:def OPEs V b} and \eqref{eq:def OPEs V c}) we find 
\begin{multline} 
\left\langle J^3(z) \prod_{i=1}^n V_{h_i}^{w_i}(x_i;z_i) \right \rangle = -\sum_{i=1}^n \frac{\mathcal{D}_i^3}{z-z_i} \left\langle \prod_{l=1}^n V_{h_l}^{w_l}(x_l;z_l) \right \rangle \\
 + \sum_{i=1}^n \sum_{\ell=1}^{w_i} \frac{x_i}{(z-z_i)^{\ell+1}} \left\langle (J^+_\ell V_{h_i}^{w_i})(x_i;z_i) \prod_{l\ne i}^n V_{h_l}^{w_l}(x_l;z_l) \right \rangle  \ , \label{eq:J3V correlator}
\end{multline}
and
\begin{multline} 
\left\langle J^-(z) \prod_{i=1}^n V_{h_i}^{w_i}(x_i;z_i) \right \rangle =-\sum_{i=1}^n \frac{\mathcal{D}_i^-}{z-z_i}  \left\langle \prod_{l=1}^n V_{h_l}^{w_l}(x_l;z_l) \right \rangle \\
+\sum_{i=1}^n \sum_{\ell=1}^{w_i} \frac{x_i^2}{(z-z_i)^{\ell+1}} \left\langle (J^+_\ell V_{h_i}^{w_i})(x_i;z_i) \prod_{l\ne i}^n V_{h_l}^{w_l}(x_l;z_l) \right \rangle\ . \label{eq:J-V correlator}
\end{multline}
Here the sum over $\ell$ starts at $\ell=1$, and we have used that the zero mode term ($\ell=0$) combines with the zero mode term of $J^3_0$ to produce exactly the differential operators ${\cal D}^3$ and ${\cal D}^-$, respectively. (Note that we could have also written the terms in the first line of (\ref{eq:JpVn}) in terms of ${\cal D}^+$.)

The key observation is now that all of these correlators contain the same `unknown terms' $\hat{F}^i_\ell$. On the other hand, we can use the fact that $J^-(z)$ has the correct OPEs with the primary fields, see \eqref{eq:def OPEs V c}, which implies that 
\begin{align}\label{eq:J-const}
\left\langle\Bigl(J^-(z)-2x_j J^3(z)+x_j^2 J^+(z)\Bigr)\prod_{i=1}^n V_{h_i}^{w_i}(x_i;z_i) \right \rangle=\mathcal{O}((z-z_j)^{w_j-1})\ .
\end{align}
Here the combination of currents in the bracket has been chosen such that 
\be\label{3.7}
\Bigl(J^{-(x_j)}(z)-2x_j J^{3(x_j)}(z)+x_j^2 J^{+(x_j)}(z)\Bigr) = J^-(z)  \ ,
\ee
see eq.~\eqref{eq:transprop}, so that near $z=z_j$ we can apply directly \eqref{eq:def OPEs V c}.
\smallskip

Using the correlators from above, it is straightforward to write out the left-hand-side of \eqref{eq:J-const} explicitly, and one finds 
\begin{multline}
\left\langle\Bigl(J^-(z)-2x_j J^3(z)+x_j ^2 J^+(z)\Bigr)\prod_{i=1}^n V_{h_i}^{w_i}(x_i;z_i) \right \rangle \\
=  \sum_{i\neq j}\left( \frac{2 (x_i-x_j) h_i +(x_i-x_j)^2 \partial_{x_i}}{(z-z_i)} \Bigl\langle \prod_{l=1}^n V_{h_l}^{w_l}(x_l;z_l) \Bigr\rangle +   \sum_{\ell=1}^{w_i} \frac{(x_i-x_j)^2}{(z-z_i)^{\ell+1}}\hat{F}^i_\ell\right) \ , \label{eq:constex}
\end{multline}
where $\hat{F}^i_\ell$ was defined above, see eq.~(\ref{eq:hatF}).

In particular, it is clear that the left-hand-side is regular as $z \to z_j$, but we still get $w_j-1$ equations by requiring that the coefficients of the terms 
\be \label{3.11}
1\ ,\quad (z-z_j)\ ,\quad (z-z_j)^2\ ,\quad \dots\ ,\quad (z-z_j)^{w_j-2}
\ee
vanish. This leads to $\sum_{j=1}^n(w_j-1)$ constraints, and we can try to solve them in terms of the $\sum_{i=1}^n w_i$ unknowns $\hat{F}^i_\ell$. 
While we have not managed to solve these conditions in closed form  --- and in fact, a solution only exists under certain conditions, see the discussion in the following section --- we can evaluate them for any choice of $n$-point function (and any choice of the $w_i$), and this is what we shall do below. 

\subsection{The fusion rules}\label{sec:fusion}

Before we continue, it is worth pausing to analyse these constraint equations for the simple case of a $3$-point function for generic $w_j$, $j=1,2,3$. We have found experimentally that the constraint equations \eqref{eq:J-const} generically only have a solution provided that\footnote{The constraint equations can be thought of as a matrix equation for the coefficients $\hat{F}^i_\ell$. This problem generically only has a solution provided that the rank of this matrix is maximal. Tim R\"othlisberger has recently shown that the matrix is indeed only of maximal rank if (\ref{eq:wconst}) holds. He has furthermore shown that if (\ref{eq:wconst}) does not hold, the only solution to the constraint equations is the trivial solution where all correlators vanish.}
\be \label{eq:wconst}
w_i+w_j\ge w_l-1 \ ,
\ee
where $(i,j,l)$ are all mutually disjoint. Thus the 
$3$-point function 
\be 
\left \langle V_{h_1}^{w_1}(x_1;z_1)V_{h_2}^{w_2}(x_2;z_2)V_{h_3}^{w_3} (x_3;z_3)\right \rangle
\ee
can only be non-zero provided that eq.~\eqref{eq:wconst} is satisfied, i.e.\ \eqref{eq:wconst} are part of the fusion rules. This result was already anticipated in \cite{Eberhardt:2018ouy}, see eq.~(6.5) of that paper;\footnote{Note that as explained there, eq.~(6.5) has to be convoluted with the $\pm 1$ shift from the direct fusion rules, so that the bounds are $|w_1 - w_2| - 1 \leq w  \leq w_1 + w_2 + 1$. This then agrees precisely with the above formula.} however, here we have directly deduced this from a Ward identity analysis on the worldsheet. 

One would similarly expect that the constraint equations \eqref{eq:J-const} for a generic $n$-point function will only have a solution provided that 
\be\label{eq:wconstgen}
\sum_{i\neq j} w_i \geq w_j-1
\ee
for all $j$. We have tested this extensively in explicit examples and it does seem to be true, but we do not have an analytic proof for it.\footnote{Tim R\"othlisberger has again shown abstractly that eq.~(\ref{eq:wconstgen}) is a necessary condition for the relevant matrix to be of maximal rank. However, it is not yet clear (although this is what we expect) whether the trivial solution is the only solution if (\ref{eq:wconstgen}) does not hold.}

\subsection{The recursion relations}\label{sec:recursion}

We can actually do a little better since we can say something about those terms that correspond to the zero mode action before spectral flow. In particular, it follows from  eq.~(\ref{eq:J+zero}) that 
\be \label{eq:J+shift}
(J^+_{w_i}V_{h_i}^{w_i})(x_i;z_i)=\left(h_i-\frac{(k+2) w_i}{2}+j_i\right)V_{h_i+1}^{w_i}(x_i;z_i)\ ,
\ee
where $j_i$ is the $\mathfrak{sl}(2,\mathds{R})$ spin before spectral flow, and we have used that for the $w$-spectrally flowed representation, $h$ is related to $m$ via \eqref{eq:hm relation}. As a consequence, one of the `unknown' terms can be written as 
\be\label{3.16}
\hat{F}^i_{w_i} = \left(h_i-\frac{(k+2) w_i}{2}+j_i\right) \left\langle V_{h_i+1}^{w_i}(x_i;z_i) \, \prod_{l\neq i} V_{h_l}^{w_l}(x_l;z_l) \right\rangle \ ,
\ee
i.e.\ as a correlator of spectrally flowed highest weight states but with a shifted value of $h_j$. By the same token we also know that 
 \be \label{eq:J-shift}
(J^-_{-w_j}V_{h_j}^{w_j})(x_j;z_j)=\left(h_j-\frac{(k+2) w_j}{2}-j_j\right)V_{h_j-1}^{w_j}(x_j;z_j)\ ,
\ee
as follows from eq.~(\ref{eq:J-zero}). Actually, because of \eqref{3.7}, we can obtain its correlator by considering the term of order $(z-z_j)^{w_j-1}$ in \eqref{eq:J-const}, i.e.\ we have 
\begin{multline}
 \left\langle \Bigl(J^-(z)-2x_j J^3(z)+x_j^2 J^+(z)\Bigr) \prod_{i=1}^n V_{h_i}^{w_i}(x_i;z_i) \right \rangle  \\
 = \left(h_j-\tfrac{(k+2)w_j}{2}-j_j\right) \, (z-z_j)^{w_j-1}\left\langle V_{h_j-1}^{w_j}(x_j;z_j)\prod_{i\ne j}^n V_{h_i}^{w_i}(x_i;z_i) \right \rangle \\
 +\mathcal{O}((z-z_j)^{w_j})\ , \label{eq:J-zeroa} 
\end{multline}
for $z$ near each $z_j$. Using Cauchy's formula we can extract the leading term on the right-hand-side as  
\be \label{3.19}
\oint_{z_j} \frac{\mathrm{d}z}{(z-z_j)^{w_j}} \left\langle \Bigl(J^-(z)-2x_j J^3(z)+x_j^2 J^+(z)\Bigr) \prod_{i=1}^n V_{h_i}^{w_i}(x_i;z_i) \right \rangle\ .
\ee
On the other hand, eq.~(\ref{3.19}) can now be evaluated in terms of the correlators from above, see eqs.~(\ref{eq:JpVn})--(\ref{eq:J-V correlator}), which however again depend on the $\hat{F}^i_\ell$.
\smallskip

Recall that the constraint equations (\ref{eq:J-const}) lead to $\sum_{j=1}^{n} (w_j-1)$ conditions on the $\hat{F}^i_\ell$, see the discussion below eq.~(\ref{3.11}). Requiring that eq.~(\ref{3.19}) has to agree with the first term in the second line of eq.~(\ref{eq:J-zeroa}) gives us now $n$ additional relations. Thus we have altogether $\sum_{j=1}^{n} w_j$ equations for the $\sum_{j=1}^{n} w_j$ unknowns $\hat{F}^i_\ell$. Thus if we can eliminate the `truly unknown' coefficients $\hat{F}^i_\ell$ with $1\leq \ell \leq w_i-1$ --- recall that $\hat{F}^i_{w_i}$ can be expressed as (\ref{3.16}) --- then we get $n$ ``recursion relations" involving only correlation functions of spectrally flowed affine highest weight states (albeit with in general shifted values of $h_i$). We have checked experimentally that $n$ linearly independent recursion relations can be derived provided that the constraint equations have a solution, i.e.\ provided that eq.~\eqref{eq:wconstgen} is satisfied.\footnote{We thank Tim R\"othlisberger for pointing out an error in a previous version of the manuscript.}
For instance, when $w_i=1$ for all $i$ --- in this case eq.~\eqref{eq:wconstgen} is always satisfied --- these ``recursion relations" read
\begin{align}
&\left(h_j-\tfrac{k+2}{2}-j_j\right) \Bigl\langle V_{h_j-1}^{1}(x_j;z_j)\prod_{\ell\ne j}^n V_{h_\ell}^{1}(x_\ell;z_\ell) \Bigr\rangle\nonumber\\
&\qquad=\sum_{i \ne j} \Bigg[\left(h_i-\tfrac{k+2}{2}+j_i\right)\frac{(x_j-x_i)^2}{(z_j-z_i)^2} \Bigl\langle V_{h_i+1}^{1}(x_j;z_j)\prod_{\ell\ne i}^n V_{h_\ell}^{1}(x_\ell;z_\ell) \Bigr\rangle\nonumber\\
&\qquad\qquad\qquad\qquad-\frac{2h_i(x_i-x_j)+(x_i-x_j)^2 \partial_{x_i}}{(z_i-z_j)} \Bigl\langle \prod_{\ell=1}^n V_{h_\ell}^{1}(x_\ell;z_\ell) \Bigr\rangle\Bigg]\ .\label{eq:wi=1 recursion relations}
\end{align}
At this point, we find it useful to make use of the global Ward identities of the theory (i.e.\ the M\"obius symmetry that acts on the $z_i$, as well as the one acting on the $x_i$), which allows us to set $x_1=z_1=0$, $x_2=z_2=1$ and $x_3=z_3=\infty$.\footnote{Sending $x_3$ and $z_3$ to infinity requires, as usual, some care in properly normalising the vertex operators. Also the derivatives in \eqref{eq:wi=1 recursion relations} on the three special fields have to be evaluated carefully using global Ward identities.} As an illustration eq.~\eqref{eq:wi=1 recursion relations} for $j=1$, say, becomes after these steps (we have dropped the dependence of the vertex operators on $(x_i;z_i)$ to make the formula more readable)
\begin{align}
&\left(h_1-\tfrac{k+2}{2}-j_1\right) \Bigl\langle V_{h_1-1}^1\, V_{h_2}^1\, V_{h_3}^1\prod_{i=4}^n V_{h_i}^{1} \Bigr\rangle \nonumber\\
&\ \ =\Biggl[h_1-h_2-h_3+\sum_{i=4}^n \frac{h_i(z_i-2x_i)+x_i(z_i-x_i) \partial_{x_i}}{z_i}\Biggr] \Bigl\langle V_{h_1}^1\,V_{h_2}^1\,V_{h_3}^1\prod_{i=4}^n V_{h_i}^{1} \Bigr\rangle \nonumber\\
&\qquad+\sum_{i=2}^n \left(h_i-\tfrac{k+2}{2}+j_i\right)\frac{x_i^2}{z_i^2} \Bigl\langle V_{h_1}^1\,V_{h_2}^1\, V_{h_3}^1V_{h_i+1}^{1} \!\!\!\! \prod_{\ell\ne \{1,2,3,i\}}^n \!\!\!\! V_{h_\ell}^{1} \Bigr\rangle\ , \label{eq:wi=1 recursion relations fixed}
\end{align}
with the convention $\frac{0}{0}=\frac{\infty}{\infty}=1$.
The recursion relations for higher values of $w_i$'s can be found accordingly, and we have implemented this in \texttt{Mathematica}. We have found, again experimentally, that these $n$ recursion relations are always mutually compatible and hence can be simultaneously solved.

In the following we shall actually assume a slightly stronger condition than eq.~(\ref{eq:wconstgen}), namely 
\be\label{eq:wconstgen2}
\sum_{i\neq j} (w_i-1) \geq w_j-1 \ , \qquad \hbox{for all $j$.}
\ee
This is a necessary condition for a covering map with the respective ramification indices to exist, and we will only be able to find nice solutions in this case. We suspect that the solutions to the recursion relations for the casese where only eq.~(\ref{eq:wconstgen}) is satisfied (but not eq.~(\ref{eq:wconstgen2})), are somewhat pathological, but we do not understand this issue in detail at present.

\subsection{Comparison to Maldacena-Ooguri}

Before we proceed, let us make another cross-check of our analysis by applying the above arguments to the case of the $3$-point function involving two unflowed representations ($w_1=w_2=0$) and one spectrally flowed representation ($w_3=1$) --- this correlator was determined by some other method in \cite{Maldacena:2001km}, see eq.~(5.38) of that paper. In this case we find from \eqref{eq:JpVn} together with \eqref{eq:J+shift} 
\begin{align} 
& \langle J^+(z) V^0_{h_1}(x_1;z_1) \, V^0_{h_2}(x_2;z_2)\, V^1_{h_3}(x_3;z_3) \rangle  \nonumber\\  
& \qquad = \sum_{i=1}^{3} \frac{1}{(z-z_i)} \partial_{x_i}  \langle  V^0_{h_1}(x_1;z_1) \, V^0_{h_2}(x_2;z_2)\, V^1_{h_3}(x_3;z_3) \rangle  \nonumber\\ 
& \qquad \qquad + \frac{1}{(z-z_3)^2} \, \bigl(h_3 - \tfrac{\hat{k}}{2} + j_3\bigr) \, \langle  V^0_{h_1}(x_1;z_1) \, V^0_{h_2}(x_2;z_2)\, V^1_{h_3+1}(x_3;z_3) \rangle \ ,
\end{align} 
where $\hat{k}=k+2$ is the level that appears in the bosonic analysis of \cite{Maldacena:2001km}. Similarly, we obtain for 
the $J^3(z)$ and the $J^-(z)$ correlators, see eqs.~\eqref{eq:J3V correlator} and \eqref{eq:J-V correlator}
\begin{align} 
& \langle J^3(z) V^0_{h_1}(x_1;z_1) \, V^0_{h_2}(x_2;z_2)\, V^1_{h_3}(x_3;z_3) \rangle \nonumber\\  
& \qquad = \sum_{i=1}^{3} \frac{1}{(z-z_i)} (h_i + x_i \partial_{x_i})  \langle  V^0_{h_1}(x_1;z_1) \, V^0_{h_2}(x_2;z_2)\, V^1_{h_3}(x_3;z_3) \rangle  \nonumber\\ 
& \qquad \qquad + \frac{x_3}{(z-z_3)^2} \, \bigl(h_3 - \tfrac{\hat{k}}{2} + j_3\bigr) \, \langle  V^0_{h_1}(x_1;z_1) \, V^0_{h_2}(x_2;z_2)\, V^1_{h_3+1}(x_3;z_3) \rangle \ ,
\end{align}
and 
\begin{align} 
& \langle J^-(z) V^0_{h_1}(x_1;z_1) \, V^0_{h_2}(x_2;z_2)\, V^1_{h_3}(x_3;z_3) \rangle \nonumber\\ 
& \qquad = \Bigl( \sum_{i=1}^{3} \frac{1}{(z-z_i)} (2 h_i x_i + x_i^2 \partial_{x_i}) \Bigr) \langle  V^0_{h_1}(x_1;z_1) \, V^0_{h_2}(x_2;z_2)\, V^1_{h_3}(x_3;z_3) \rangle  \nonumber\\ 
&  \qquad \qquad + \frac{x_3^2}{(z-z_3)^2} \, \bigl(h_3 - \tfrac{\hat{k}}{2} + j_3\bigr) \, \langle  V^0_{h_1}(x_1;z_1) \, V^0_{h_2}(x_2;z_2)\, V^1_{h_3+1}(x_3;z_3) \rangle \ .
\end{align}
Then imposing the condition that the regular (i.e.\ the $(z-z_3)^0$ term) of the correlator in eq.~\eqref{eq:J-zeroa} satisfies
\begin{multline}
\left. \Bigl\langle \Bigl( J^-(z) - 2 x_3 J^3(z) + x_3^2 J^+(z) \Bigr) V^0_{h_1}(x_1;z_1) \, V^0_{h_2}(x_2;z_2)\, V^1_{h_3}(x_3;z_3) \Bigr\rangle  \right|_{(z-z_3)^0} \\
 = \bigl(h_3 - \tfrac{\hat{k}}{2} - j_3\bigr) \langle  V^0_{h_1}(x_1;z_1) \, V^0_{h_2}(x_2;z_2)\, V^1_{h_3-1}(x_3;z_3) \rangle \ ,
\end{multline}
leads to the condition that 
\begin{multline}
\langle  V^0_{h_1}(x_1;z_1) \, V^0_{h_2}(x_2;z_2)\, V^1_{h_3-1}(x_3;z_3) \rangle \,  \frac{(x_1-x_2)}{(x_1-x_3) (x_2-x_3)}\\ 
= \frac{ h_1 + h_2 - h_3}{h_3 - \frac{\hat{k}}{2} - j_3} \, 
\langle  V^0_{h_1}(x_1;z_1) \, V^0_{h_2}(x_2;z_2)\, V^1_{h_3}(x_3;z_3) \rangle \ , 
\end{multline}
which is compatible with eq.~(5.38) of \cite{Maldacena:2001km}. (Here we have used that $h_1=j_3$, $h_2=j_4$ and $J=h_3$; we have also set $j_1^{\rm MO} = 1 - j_3$.) Thus our analysis is compatible with the results of \cite{Maldacena:2001km}.

\section{A simple solution} \label{sec:simple solution}

While we do not yet have a clear understanding of what the most general solution of the recursion relations is, there is a remarkably simple (and suggestive) solution that works very generally. It involves critically the covering map, whose structure we shall now review.

\subsection{The covering map}\label{sec:gencov}

Suppose $\{z_i\}_{i=1,\ldots,n}$ and $\{x_i\}_{i=1,\ldots,n}$ are two collections of $n$ points on the Riemann sphere,\footnote{In Section~\ref{sec:genus} below, we shall also comment on the case where the $x_i$ live on the Riemann sphere, but the $z_i$ may be taken to lie on any Riemann surface of genus $g$.} and $\{w_i\}_{i=1,\ldots,n}$ are positive integers, describing the ramification indices near $z_i$. We call $\Gamma(z)$ the \emph{covering map} of this configuration provided that $\Gamma(z)$ is an analytic function satisfying
\be\label{eq:coveringdef}
\Gamma(z) = x_i + a_i^\Gamma (z-z_i)^{w_i} + {\cal O}((z-z_i)^{w_i+1}) 
\ee
near $z=z_i$. In addition, we assume that $\Gamma(z)$ does not have any other critical points, i.e.\ the only points where $\partial\Gamma(z)=0$ is for $z=z_i$ with $i=1,\ldots,n$. 
\smallskip

There are a few general statements about covering maps one can make. Let us first consider the case that $n=3$. By composing the covering map with suitable M\"obius transformations, we may assume, without loss of generality, that $x_i=z_i$, $i=1,2,3$. Provided that the $w_i$ satisfy the selection rule 
\be\label{eq:selection}
w_i + w_j \geq w_k  +1 \ , \qquad w_1 + w_2 + w_3 \in 2 \mathds{Z}+1\ ,
\ee
a covering map exists, and it is unique. We should mention in passing that, up to the parity constraint that the sum of the $w_i$ must be odd, this agrees precisely with the conditions under which a solution to the recursion relations can be found, see eq.~(\ref{eq:wconstgen2}). 

For example, for $w_i=1$, the covering map is simply $\Gamma(z)=z$, while for the case $w_1=w_2=2$, $w_3=1$ it is 
\be
\Gamma(z) = \frac{z^2 ( z_1 + z_2 - z_3) - 2 z z_1 z_2 + z_1 z_2 z_3}{ z^2 - 2 z z_3 +(z_1 z_3 +z_2 z_3 - z_1z_2)} \ , 
\ee
as one can verify easily. Note that this covering map is $2$-to-$1$, i.e.\ every generic point in $\mathrm{S}^2$ has two preimages. (In particular, since the denominator is a quadratic polynomial in $z$, there are two points that are mapped to infinity.) In general the degree of the covering map, i.e.\ the number of preimages of a generic point, equals 
\be
N = 1 + \frac{1}{2} \sum_{i=1}^{n} (w_i - 1) \ , \label{eq:degree covering map}
\ee
as follows from the Riemann-Hurwitz formula. Incidentally, this result is true for any $n$ (not just $n=3$). 
\smallskip

While the covering map always exists for $n=3$ (provided that \eqref{eq:selection} is satisfied), the situation for $n\geq 4$ is significantly different: in this case, the covering map generically \emph{does not exist}. While there is a natural analogue of \eqref{eq:selection}, which in the general case takes the form\footnote{We note again the similarity of the first condition to \eqref{eq:wconstgen2}. We should also mention that both conditions in (\ref{covex}) follow from the Riemann-Hurwitz formula (\ref{eq:degree covering map}): the inequality comes from the fact that by construction $N\geq w_i$ for every $i$, while the parity constraint comes from the requirement that $N$ has to be an integer.}
\be\label{covex}
\sum_{i\neq j} (w_i-1) \geq w_j-1 \ , \qquad \sum_{i} (w_i-1) \in 2 \mathds{Z} \ ,
\ee
this is in general not enough to guarantee the existence of the covering map. In order to see this let us consider the simple case that $n=4$ with $w_i=1$. We may again use the M\"obius symmetry to arrange for $x_i=z_i$ for $i=1,2,3$, but now we do not have any freedom left to choose $x_4$ and $z_4$. Since $w_i=1$, the covering map must be $\Gamma(z)=z$, but this now maps $\Gamma(z_4)=z_4$, which only agrees with $x_4$ provided that $x_4=z_4$. Thus, in the $4$-point case, the covering map only exists provided the $x_i$ satisfy one constraint. This analysis generalises to $n\geq 4$ (and arbitrary $w_i$), and in general the $x_i$ have to satisfy $n-3$ such constraints. 

To illustrate this phenomenon, let us look at the covering map for the case $w_1=w_2=w_3=w_4=2$. Assuming (for simplicity) that $z_1=x_1=0$, $z_2=x_2=1$, as well as $z_3=x_3=\infty$, it is explicitly given by
\be\label{cov2222}
\Gamma(z) = \frac{z^2 \left(\pm z \sqrt{z_4^2- z_4+1}-z z_4-z+3 z_4\right)}{\pm (3 z-2) \sqrt{z_4^2-z_4+1}+3 z z_4-3 z-z_4+2} \ .
\ee
In this case there are actually two possible covering maps corresponding to the two possible sign choices in front of the square root. Furthermore, in either case this only defines an actual covering map provided that $x_4$ satisfies the constraint
\be 
x_4=\Gamma(z_4) = z_4 \Bigl(\pm \, 2(1-z_4) \sqrt{1 - z_4 + z_4^2} + 2 z_4^2 - 3 z_4 + 2  \Bigr) \ . 
\ee
\smallskip

Since the correlation functions will naturally involve the covering map, this structural difference between $n=3$ and $n\geq 4$ will also be reflected in the correlation functions. We shall therefore describe the relatively simple case of a $3$-point function first, before explaining how the analysis generalises to $n\geq 4$.

\subsection{The 3-point case}

Let us denote by $a_i^\Gamma$, $i=1,2,3$, the coefficients that appear in the covering map as in \eqref{eq:coveringdef}. We claim that a solution to the above recursion relations is given by setting 
\begin{multline}
 \left \langle V_{h_1}^{w_1}(x_1;z_1)V_{h_2}^{w_2}(x_2;z_2)V_{h_3}^{w_3} (x_3;z_3)\right \rangle \\
 =C(j_1,j_2,j_3)\prod_{i=1}^3 (a_i^\Gamma)^{-h_i}
\prod_{i\neq j} (z_i - z_j)^{\Delta_{\ell}^0- \Delta^0_i- \Delta^0_j} \ ,\label{eq:sol 3-point function}
\end{multline}
where in the above product $\ell$ labels the third index, not equal to either $i$ or $j$. 
The overall coefficient $C(j_1,j_2,j_3)$ is just a normalisation constant, and we have assumed that the $j_i$ satisfy the relation 
\be
\sum_{i=1}^{3} j_i = \frac{k+2}{2} \ , \label{eq:j constraint 3-point function}
\ee
where $k$ is the level of the supersymmetric $\mathfrak{sl}(2,\mathbb{R})_{k}^{(1)}$ affine algebra. (The decoupled bosonic algebra has then level $\hat{k}=k+2$.) In addition, we have set 
\be
\Delta_j^0 = \Delta_j + w_j h_j \ , 
\ee
where $\Delta_j$ is the conformal dimension of the corresponding vertex operator, see eq.~(\ref{Deltadef}), although this does not play a role in the following. (The $z_i$ dependence is largely irrelevant for our analysis.) 

We should note that the expression in (\ref{eq:sol 3-point function}) depends implicitly also on the $x_i$'s, since the coefficients $a_i^\Gamma$ of the covering map depend on all $x_i$'s as well as the $z_i$'s. In fact, we have explicitly \cite{Lunin:2000yv}
\be 
a_i^\Gamma=\frac{\binom{\frac{1}{2}(w_i+w_{i+1}+w_{i+2}-1)}{\frac{1}{2}(-w_i+w_{i+1}+w_{i+2}-1)}}{\binom{\frac{1}{2}(-w_i+w_{i+1}-w_{i+2}-1)}{\frac{1}{2}(w_i+w_{i+1}-w_{i+2}-1)}} \, \frac{(x_i-x_{i+1})(x_{i+2}-x_i)(z_{i+1}-z_{i+2})^{w_i}}{(x_{i+1}-x_{i+2})(z_i-z_{i+1})^{w_i}(z_{i+2}-z_i)^{w_{i}}} \ ,\label{eq:ai 3-point function}
\ee
where indices are understood to be $\bmod \ 3$. Hence, the $3$-point function in (\ref{eq:sol 3-point function}) has indeed the expected $x_i$ and $z_i$ dependence,
\begin{multline}
 \left \langle V_{h_1}^{w_1}(x_1;z_1)V_{h_2}^{w_2}(x_2;z_2)V_{h_3}^{w_3} (x_3;z_3)\right \rangle \\
 =C(j_1,j_2,j_3)\prod_{i=1}^3 \left(\frac{\binom{\frac{1}{2}(w_i+w_{i+1}+w_{i+2}-1)}{\frac{1}{2}(-w_i+w_{i+1}+w_{i+2}-1)}}{\binom{\frac{1}{2}(-w_i+w_{i+1}-w_{i+2}-1)}{\frac{1}{2}(w_i+w_{i+1}-w_{i+2}-1)}}\right)^{\!\!\!-h_i}\\
\times
\prod_{i\neq j} (z_i - z_j)^{\Delta_{\ell} - \Delta_i - \Delta_j}(x_i - x_j)^{h_{\ell} - h_i - h_j} \ .
\end{multline}
We shall show in the following subsection (Section~\ref{sec:3pointproof}) that this ansatz satisfies in fact the recursion relations of Section~\ref{sec:recursion}. We have also checked this for many explicit examples by direct computation.

It is sometimes convenient to use the M\"obius symmetry on the worldsheet and in spacetime to fix
 $z_1=x_1=0$, $z_2=x_2=1$ and $z_3=x_3=\infty$; here for $z_3=x_3=\infty$, the relevant condition on the covering map is\footnote{This can be seen by a coordinate transformation $z \to -\tfrac{1}{z}$ and $x \to -\tfrac{1}{x}$.}
\be 
\Gamma(z)=(-1)^{w_3+1} \, \frac{z^{w_3}}{a_3^\Gamma}+\mathcal{O}(z^{w_3-1})\ .
\ee
We then simply have
\be \label{eq:3point}
 \left \langle V_{h_1}^{w_1}(0;0)V_{h_2}^{w_2}(1;1)V_{h_3}^{w_3} (\infty;\infty)\right \rangle=C(j_1,j_2,j_3)\prod_{i=1}^3  (a_i^\Gamma)^{-h_i}\ ,
\ee
where $a_i^\Gamma$ is given by the ($x_i$ and $z_i$ independent) ratio of binomial coefficients that appears in the first factor of \eqref{eq:ai 3-point function}.

\subsection{The proof of the 3-point solution}\label{sec:3pointproof}

In this section we want to prove that the ansatz \eqref{eq:sol 3-point function} satisfies indeed all the recursion relations of Section~\ref{sec:recursion}. In order to treat the  insertion points uniformly, we assume here that $x_i$ and $z_i$ are generic.

Our basic strategy is to first show that the constraint equations of Section~\ref{sec:basic strategy} can be reformulated as a functional identity (see eq.~(\ref{eq:G propto dGamma})) for a function whose definition involves the covering map, see eq.~(\ref{eq:G definition}). Implicitly, this therefore solves the constraint relations, and allows us to write the recursion equations as the equality of eqs.~(\ref{eq:left hand side}) and (\ref{eq:right hand side}). Finally, we show that our ansatz \eqref{eq:sol 3-point function} satisfies indeed these relations.
\smallskip

We start by rewriting the constraint equation \eqref{eq:J-const} by dividing both sides of \eqref{eq:constex} by the correlator. Then the constraint equations are equivalent to 
\begin{align}
0= &\left. \frac{\left\langle\Bigl(J^-(z)-2x_j J^3(z)+x_j^2 J^+(z)\Bigr)\prod_{i=1}^3 V_{h_i}^{w_i}(x_i;z_i) \right \rangle }{\left\langle\prod_{l=1}^{3} V_{h_l}^{w_l}(x_l;z_l) \right \rangle }\right|_{(z-z_j)^m}\nonumber\\
&\qquad \qquad 
=\frac{1}{m!} \partial_{z}^m\left.\sum_{i \ne j}\left(-\frac{2(x_j-x_i)h_j}{z-z_i}+\sum_{\ell=0}^{w_i} \frac{(x_j-x_i)^2}{(z-z_i)^{\ell+1}} F_\ell^i\right)\right|_{z=z_j} \ , \label{eq:rewriting Jm constraint}
\end{align}
where $m=0,1,\ldots,w_j-2$ and $j=1,\ldots,n$. We have also defined 
\be\label{eq:Fdef}
F^i_\ell = \frac{\hat{F}^i_\ell}{\Bigl\langle \prod_{l=1}^{3} V_{h_l}^{w_l}(x_l;z_l) \Bigr\rangle}  
= \frac{\Bigl\langle (J^+_\ell V^{w_i}_{h_i}) (x_i;z_i) \, \prod_{l\neq i} V_{h_l}^{w_l}(x_l;z_l) \Bigr\rangle}{\Bigl\langle \prod_{l=1}^{3} V_{h_l}^{w_l}(x_l;z_l) \Bigr\rangle} \ ,
\ee
see eq.~\eqref{eq:hatF}. (Here the derivative term of \eqref{eq:constex} is included in the sum as $\ell=0$.)
If we denote by $\Gamma(z)$ the relevant covering map, we can rewrite the right-hand-side of \eqref{eq:rewriting Jm constraint} as 
\be\label{4.18}
\frac{1}{m!} \partial_{z}^m\left.\sum_{i=1}^{3}\left(-\frac{2(\Gamma(z)-x_i)h_j}{z-z_i}+\sum_{\ell=0}^{w_i} \frac{(\Gamma(z)-x_i)^2}{(z-z_i)^{\ell+1}} F_\ell^i\right)\right|_{z=z_j} \ , 
\ee
where we have first replaced $x_j=\Gamma(z_j)$, and then extended the sum also over $i=j$; the first operation will only affect the result at order $m=w_j$, while extending the sum to $i=j$ will also have an affect at order $m=w_j-1$, see below. Thus, defining the function 
\be 
G(z)=\sum_{i=1}^3\left(-\frac{2(\Gamma(z)-x_i)h_i}{z-z_i}+\sum_{\ell=0}^{w_i} \frac{(\Gamma(z)-x_i)^2}{(z-z_i)^{\ell+1}} F_\ell^i\right)\ ,\label{eq:G definition}
\ee
the constraint equations on the $F_\ell^i$ are equivalent to the requirement that 
\be 
\partial_z^m G(z=z_j)=0
\ee
for $m=0,\dots,w_j-2$ and $j=1,\ldots,n$. Assuming that we have solved these constraints, $G(z)$ is a rational function with the following properties:
\begin{enumerate}
\item $G(z)$ has double poles at $\{z_a^*\}$, $a=1,\dots,N$, where $z_a^*$ are the poles of $\Gamma(z)$. Here $N$ is given by the Riemann-Hurwitz formula, see eq.~\eqref{eq:degree covering map}.
\item $G(z)$ has zeros of order $w_i-1$ at the insertion points.
\item $G(z)$ behaves asymptotically as $\mathcal{O}(z^{-2})$.\label{prop:3}
\end{enumerate}
The last property requires some explanation. The asymptotic expansion of $G(z)$ reads
\begin{align} 
G(z)
&=\frac{1}{z}\sum_{i=1}^3\Bigl(\Gamma(\infty)^2 F_0^i-2\Gamma(\infty)(h_i+x_i F_0^i)+(2x_i h_i+x_i^2 F_0^i)\Bigr)+\mathcal{O}(z^{-2})\ .\label{eq:G expansion}
\end{align}
Global $\mathrm{SL}(2,\mathds{R})$ invariance, on the other hand, implies the Ward identity
\be \label{5.22}
0=\sum_{i=1}^3 \partial_{x_i} \langle V_{h_1}^{w_1}(x_1;z_1)V_{h_2}^{w_2}(x_2;z_2)V_{h_3}^{w_3}(x_3;z_3)\rangle=\sum_{i=1}^3 \hat{F}^i_0\ ,
\ee
and similarly for the zero modes $J^3_0$ and $J^-_0$; this leads to\footnote{In going from  (\ref{5.22}) to (\ref{eq:zero modes Ward identities}) we have divided by the correlator and used the definition (\ref{eq:Fdef}).}
\begin{align}
\sum_{i=1}^3 F_0^i&=0\ ,& \sum_{i=1}^3 (h_i+x_i F_0^i)&=0\ ,\quad\text{and} & \sum_{i=1}^3 (2x_i h_i+x_i^2 F_0^i)&=0\ , \label{eq:zero modes Ward identities}
\end{align}
which implies the vanishing of the $\frac{1}{z}$ term in \eqref{eq:G expansion}. 
\smallskip

The three properties above now imply that $G(z)$ must be of the form 
\be 
G(z)=A \ \frac{\prod_{i=1}^3(z-z_i)^{w_i-1}}{\prod_{a=1}^N (z-z_a^*)^2} \ ,
\ee
where $A$ is a normalisation constant, and we have used that 
\be\label{4.25}
\sum_{i=1}^{3} (w_i-1) - 2 N = -2 \ , 
\ee
as follows from eq.~\eqref{eq:degree covering map}. (In particular, there cannot be any additional powers of $z$ in the numerator because of property~\ref{prop:3}.)  The key observation is now that $G(z)$ therefore has exactly the same properties as the derivative of the covering map, and hence that 
\be 
G(z)=\alpha \, \partial \Gamma(z) \ ,
\ee
where $\alpha$ is some constant.\footnote{The constant $\alpha$ still depends on the remaining variables $x_i$, $z_i$, as well as $j_i$ and $k$.} 
We can compute the proportionality constant $\alpha$ by summing over all poles of the covering map --- there are always $N$ such poles, where $N$ is determined by the Riemann-Hurwitz formula (\ref{eq:degree covering map}), and we denote them by $z_a^*$ with $a=1,\ldots, N$ 
\begin{align}
\alpha N&=-\sum_{a=1}^N \oint_{z_a^*} \mathrm{d}z\ \frac{G(z)}{\Gamma(z)} \label{eq:alpha derivation first}\\
&=-\sum_{a=1}^N \sum_{i=1}^3 \oint_{z_a^*} \mathrm{d}z\ \left(-\frac{2(\Gamma(z)-x_i)h_i}{(z-z_i)\Gamma(z)}+\sum_{\ell=0}^{w_i} \frac{(\Gamma(z)-x_i)^2}{(z-z_i)^{\ell+1}\Gamma(z)} F_\ell^i\right) \\
&=-\sum_{a=1}^N \sum_{i=1}^3 \sum_{\ell=0}^{w_i}\oint_{z_a^*} \mathrm{d}z\  \frac{\Gamma(z)}{(z-z_i)^{\ell+1}} F_\ell^i \\
&=\sum_{i=1}^3 \sum_{\ell=0}^{w_i}\left(\oint_{z_{i}} \mathrm{d}z\  \frac{\Gamma(z)}{(z-z_i)^{\ell+1}} F_\ell^i+\oint_{\infty} \mathrm{d}z\  \frac{\Gamma(z)}{(z-z_i)^{\ell+1}} F_\ell^i\right) \\
&=\sum_{i=1}^3 \sum_{\ell=0}^{w_i}\left(\oint_{z_{j}} \mathrm{d}z\  \frac{x_i+a_i^\Gamma(z-z_i)^{w_i}+\cdots}{(z-z_i)^{\ell+1}} F_\ell^i-\delta_{\ell,0} \, \Gamma(\infty) F_0^i\right) \label{penultimate} \\
&=\sum_{i=1}^3 \Bigl( ( x_i - \Gamma(\infty) )\, F_0^i  + a_i^\Gamma F^i_{w_i} \Bigr) 
= \sum_{i=1}^3\left(a_i^\Gamma F_{w_i}^i-h_i\right) \ . \label{eq:alpha derivation last}
\end{align}
Here we have used the first and second Ward identity of \eqref{eq:zero modes Ward identities} in the last step. We therefore conclude that the constraint equations \eqref{eq:J-const} are equivalent to requiring that $G(z)$, as defined by (\ref{eq:G definition}), equals 
\be 
G(z)=\frac{2}{w_1+w_2+w_3-1}\sum_{i=1}^3 \left(a_i^\Gamma F_{w_i}^i-h_i\right) \partial \Gamma(z)\ .\label{eq:G propto dGamma}
\ee
(Here we have written $N=w_1+w_2+w_3-1$, see eq.~(\ref{4.25}).) In particular, by equating (\ref{eq:G definition}) with (\ref{eq:G propto dGamma}),
this allows us to determine the $F^i_\ell$ (with $0\leq \ell\leq w_i-1$) in terms of the $h_i$, $F^i_{w_i}$, as well as the covering map $\Gamma(z)$. This solves at least in principle the constraint equations.
\smallskip

Next we turn to the recursion relations of Section~\ref{sec:recursion}. Expanding $G(z)$ to order $w_j-1$ around $z_j$ leads to 
\begin{align}
\frac{\partial_z^{w_j-1} G(z=z_j)}{(w_j-1)!}&=\frac{ \partial_{z}^{w_j-1}}{(w_j-1)!}\left.\sum_{i \ne j}\left(-\frac{2(x_j-x_i)h_j}{z-z_i}+\sum_{\ell=0}^{w_i} \frac{(x_j-x_i)^2}{(z-z_i)^{\ell+1}} F_\ell^i\right)\right|_{z=z_j} \nonumber\\
&\qquad +  \bigl( -2a_j^\Gamma h_j+(a_j^\Gamma)^2 F^{j}_{w_j} \bigr) \label{eq:recursion relation derivation 1}\ , 
\end{align}
where the two terms in the second line come from the contribution with $i=j$, see the comment below eq.~(\ref{4.18}). Using eqs.~\eqref{eq:J-zeroa} and \eqref{3.16}, this leads to 
\begin{multline}
\frac{\partial_z^{w_j-1} G(z=z_j)}{(w_j-1)!}
=\left(h_j-\tfrac{(k+2)w_j}{2}-j_j\right) \, \frac{\left\langle V_{h_j-1}^{w_j}(x_j;z_j)\prod_{i\ne j}^3 V_{h_i}^{w_i}(x_i;z_i) \right \rangle}{\left\langle \prod_{i=1}^3 V_{h_i}^{w_i}(x_i;z_i) \right \rangle}-2 a_j^\Gamma h_j\\
+(a_j^\Gamma)^2 \left(h_j-\tfrac{(k+2)w_j}{2}+j_j\right) \, \frac{\left\langle V_{h_j+1}^{w_j}(x_j;z_j)\prod_{i\ne j}^3 V_{h_i}^{w_i}(x_i;z_i) \right \rangle}{\left\langle \prod_{i=1}^3 V_{h_i}^{w_i}(x_i;z_i) \right \rangle}\ , \label{eq:left hand side}
\end{multline}
where $G(z)$ is the function that was determined before in \eqref{eq:G propto dGamma}, i.e.\
\be 
\frac{\partial_z^{w_j-1} G(z=z_j)}{(w_j-1)!}=\frac{2w_j a_j^\Gamma}{w_1+w_2+w_3-1}\sum_{i=1}^3 \left(a_i^\Gamma F_{w_i}^i-h_i\right) \ .\label{eq:right hand side}
\ee
The equality of the right-hand-sides of eqs.~(\ref{eq:left hand side}) and (\ref{eq:right hand side}) is therefore the recursion relation for the case of $3$-point functions. Note that here we have used the constraint equations to eliminate the `truly unknown' coefficients. 
\medskip

It now only remains to show that our ansatz \eqref{eq:sol 3-point function} satisfies these conditions. Inserting \eqref{eq:sol 3-point function} into \eqref{eq:left hand side}, we obtain
\be 
\frac{\partial_z^{w_j-1} G(z=z_j)}{(w_j-1)!}=- (k+2) a_j^\Gamma w_j
\ee
while inserting \eqref{eq:sol 3-point function} into \eqref{eq:right hand side} leads to 
\be 
\frac{\partial_z^{w_j-1} G(z=z_j)}{(w_j-1)!}=\frac{2a_j^\Gamma w_j}{w_1+w_2+w_3-1}\sum_{i=1}^3 \left(j_i-\frac{(k+2) w_i}{2}\right)\ .
\ee
These two expressions are precisely equal to one another provided that the constraint \eqref{eq:j constraint 3-point function} holds. We have therefore shown that our ansatz  \eqref{eq:sol 3-point function} solves the recursion relations.

\subsection{The solution in the general case}\label{sec:gencorr}

For the general case of an $n$-point function, we claim that the recursion relations are solved by 
\begin{multline}
\Bigl\langle V_{h_1}^{w_1}(0;0)V_{h_2}^{w_2}(1;1)V_{h_3}^{w_3} (\infty;\infty)\prod_{i=4}^n V_{h_i}^{w_i}(x_i;z_i) \Bigr\rangle \\
=\sum_{\Gamma} \prod_{i=1}^n (a_i^{\Gamma})^{-h_i} \prod_{i=4}^n \delta(x_i-\Gamma(z_i)) \, W_\Gamma(z_4,\dots, z_n)\ . \label{eq:localisation solution}
\end{multline}
Here, each $W_\Gamma$ is an arbitrary function depending on the remaining cross-ratios on the covering sphere (as well as on all $j_i$'s and $k$). We will see that this is a solution of the recursion relations, provided that
\be 
\boxed{\sum_{i=1}^n j_i=\frac{(k+2)}{2}(n-2)-(n-3)\ .}  \label{eq:j condition}
\ee
(This is the generalisation of eq.~(\ref{eq:j constraint 3-point function}) to $n\geq 4$.) For an $n$-point function, there are typically several possible covering maps $\Gamma(z)$, see e.g.\ the example of eq.~(\ref{cov2222}), and we should naturally consider the sum over all of them. 

Before we proceed we need to explain what precisely is meant by $\Gamma$ in \eqref{eq:localisation solution}. As was mentioned above, the covering map generically \emph{does not} exist for $n\geq 4$. Here and in the following we will always take $\Gamma(z)$ to be the map that has ramification points of order $w_i$ at $z_i$, and satisfies
\be 
\Gamma(z_i)=x_i\ , \qquad i=1,\, 2,\, 3\ .
\ee
Such a function $\Gamma(z)$ always exists, provided that \eqref{covex} is satisfied (but it may not be unique, see e.g.\ the example in eq.~(\ref{cov2222})). It defines a relevant 
covering map if $x_i=\Gamma(z_i)$ for $i=4,\,\dots,\, n$, i.e.~if the arguments of the delta-functions in \eqref{eq:localisation solution} vanish. 
\smallskip

For illustration, we can test eq.~(\ref{eq:localisation solution}) for the case when all $w_i=1$, for which we have given the recursion relations explicitly in eq.~\eqref{eq:wi=1 recursion relations fixed}. In this case, the unique map $\Gamma$ is $\Gamma(z)=z$, and hence $a_i^\Gamma=1$ for all $i$. Using the distributional identity
\be 
x \delta'(x)=-\delta(x)\ ,
\ee
eq.~\eqref{eq:wi=1 recursion relations fixed} becomes
\begin{multline}
\left(h_1-\tfrac{k+2}{2}-j_1\right) \, \prod_{i=4}^n \delta(x_i-z_i) \, W_\Gamma(z_4,\dots, z_n) \\
= \Bigl(\sum_{i=2}^n \left(h_i-\tfrac{k+2}{2}+j_i\right)+(n-3)+ h_1-\sum_{i=2}^n h_i \Bigr) \prod_{i=4}^n \delta(x_i- z_i) \, W_\Gamma(z_4,\dots, z_n)\ ,
\end{multline}
which holds indeed provided that \eqref{eq:j condition} is obeyed. We should note that the correction $n-3$ on the right-hand-side of \eqref{eq:j condition} is produced by the presence of the $(n-3)$ $\delta(x_i-z_i)$ functions. 
We have also checked explicitly that eq.~(\ref{eq:localisation solution}) satisfies the recursion relations for some other cases with small values of $w_i$. 

One can also generalise the argument of Section~\ref{sec:3pointproof} to the general case and prove that this is always a solution to the recursion relations. Since, the presence of the $\delta$-functions requires extra care, this is described in Appendix~\ref{app:npointsol}.

\subsection{The \texorpdfstring{$j$}{j} condition}\label{sec:jcond}

In the previous subsections we have seen that a solution for these correlation functions can be found provided that eq.~\eqref{eq:j condition} is satisfied. One may wonder what the meaning of this condition on the $j_i$ is (and how one may satisfy it in general). A natural solution that works for general $n$ and $k$ is given by setting\footnote{Alternatively, one could also modify the condition on the $j$'s of eq.~\eqref{eq:j condition} by sending some $j_i \to 1-j_i$. Then there also exists a natural solution to the recursion relations, for which the factor of $a_i^\Gamma$ in (\ref{eq:localisation solution}) is modified by some additional binomial coefficients. However, here we shall always work with the simple solution that arises if we impose \eqref{eq:j condition}.} 
\be \label{jsol}
j_1=1-\frac{k}{2}\ , \quad \qquad  j_i=\frac{k}{2} \ ,\quad i\geq 2 \ . 
\ee 
The value of $ j=\frac{k}{2}$ describes precisely the worldsheet representation that corresponds to the twisted sector ground states in the symmetric product orbifold \cite{Dei:2019osr}, while $j=1-\frac{k}{2}$ is the conjugate representation. 
\smallskip

Actually, the situation is particularly clean for the case $k=1$. Recall that it was recently shown in \cite{Eberhardt:2018ouy}, see also \cite{Gaberdiel:2018rqv}, that the spectrum of superstring theory on $\text{AdS}_3 \times \text{S}^3 \times \mathbb{T}^4$ at $k=1$ agrees exactly with that of the symmetric product orbifold of $\mathbb{T}^4$. A key step in making this precise was to note that at $k=1$ the {\it only} allowed worldsheet representations have $j=\tfrac{1}{2}$ \cite{Eberhardt:2018ouy}. Our main observation now is that 
\begin{verse}
{\it Eq.~\eqref{eq:j condition} is always solved for $j_i=\frac{1}{2}$ and $k=1$.}
\end{verse}
(Another way of saying this is that all $j_i$ in (\ref{jsol}) are actually equal to $j_i=\frac{1}{2}$ for $k=1$.)

In particular, our arguments therefore show that the corresponding correlation functions for the theory that is dual to the symmetric orbifold \emph{always} localise. This ties in very nicely with the expectations from the dual symmetric orbifold. It also suggests strongly that this part of the $k=1$ theory is in some sense topological, as was already noted based on the structure of the torus partition function in \cite{Eberhardt:2018ouy}.

\section{The covering map as a correlator}\label{sec:covering}

As we have seen above, the covering map plays an important role in the construction of the solution to the recursion relations. However, much more is actually true: one can also obtain the covering map directly as a correlator, provided one considers the Wakimoto representation of the $\mathfrak{sl}(2,\mathds{R})_{k+2}$ current algebra.

\subsection{The Wakimoto representation}

The Wakimoto representation of the $\mathfrak{sl}(2,\mathds{R})_{k+2}$ current algebra is given by  \cite{Giveon:1998ns, Wakimoto:1986gf}
\begin{subequations}\label{eq:Wakimoto}
\begin{align}
J^+&=k \beta\ , \\
J^3&=-k \partial \Phi+k (\beta\gamma)\ , \\
J^-&=-2 k (\partial \Phi \gamma)+k (\beta\gamma\gamma)- (k+2) \partial \gamma\ ,
\end{align}
\end{subequations}
where $\beta(z)$, $\gamma(z)$ and $\partial \Phi(z)$ are free fields with defining OPEs
\be 
\beta(z) \gamma(w) \sim -\frac{1}{k(z-w)}\ ,\qquad \partial \Phi(z) \partial \Phi(w) \sim -\frac{1}{2k(z-w)^2}\ .
\ee
The energy-momentum tensor takes the form
\be \label{Tdef}
T(z)=-k(\partial \Phi)^2-\partial^2\Phi-k (\beta \partial \gamma)\ .
\ee
Thus, we see that $\gamma(z)$ has conformal weight zero (and $\mathfrak{sl}(2,\mathds{R})$ charge $-1$) while $\beta(z)$ has conformal weight one (and $\mathfrak{sl}(2,\mathds{R})$ charge $+1$). Importantly, we also see that the boson $\Phi(z)$ has a background charge, which will lead to an anomalous conformal transformation behaviour.

\subsection{The correlator of \texorpdfstring{$\gamma$}{gamma}}

Our main claim is that, provided the condition \eqref{eq:j condition} on $j$ holds, we have the identity
\be 
\Bigl\langle \gamma(z) \prod_{i=1}^n V_{h_i}^{w_i}(x_i;z_i) \Bigr\rangle=\Gamma(z) \, \Bigl\langle \prod_{i=1}^n V_{h_i}^{w_i}(x_i;z_i) \Bigr\rangle\ . \label{eq:covering map}
\ee
In particular, $\gamma(z)$ is therefore the field, which when inserted in a correlation function, yields the corresponding covering map. This idea was already anticipated in \cite{Eberhardt:2019qcl}, but now we can make this more precise. 
\smallskip

There are several subtleties associated to this statement. First of all, the covering map $\Gamma(z)$ has poles away from the insertion points $z_i$. (There are always $N$ such poles,  where $N$ is determined by (\ref{eq:degree covering map}), and their location will  be denoted by $z_a^*$ with $a=1,\ldots, N$.) Thus, the correlator with $\gamma(z)$ inserted has to have the property that there are poles at  $z=z_a^*\neq z_j$. Initially, this does not seem to make any sense since the only field insertions are at $z=z_j$. 
The resolution to this problem is that effectively charge conservation requires that there are additional (``secret") fields present in the correlator. These secret fields behave as the vacuum with respect fo the $\mathfrak{sl}(2,\mathds{R})$ currents, but have non-trivial OPEs with the Wakimoto fields $\gamma(z)$ (and $\partial\Phi$); we show that such fields do in fact exist in Appendix~\ref{app:secret}. While this does not directly explain why these secret fields have to sit at the $z_a^*$, it shows at least that this is not in contradiction with what we have done so far (where we have only used Ward identities involving the affine currents, for which these secret fields are invisible). 

Secondly, we observe that \eqref{eq:covering map} is in general an equation between distributions since the covering map $\Gamma(z)$ only exists when the points $x_i$ and $z_i$ are suitably chosen. In particular, the correlator on the right-hand-side has delta-function support localising it to the points where the covering map exists. For a given choice of $x_i$ and $z_i$ then only one term in the sum \eqref{eq:localisation solution} can be non-zero, and the correlator picks out the corresponding covering map.
\smallskip

After these preliminary comments, let us now summarise the evidence we have for eq.~\eqref{eq:covering map}. Using the Wakimoto representation and the defining OPEs \eqref{eq:def OPEs V} and  (\ref{eq:Jxshift}), it follows that the OPE of $\gamma(z)$ with a spectrally flowed affine primary field takes the form 
\begin{align}
\gamma(z) V_{h_i}^{w_i}(x_i;z_i) \sim & \ x_i V_{h_i}^{w_i}(x_i;z_i)+(z-z_i)^{w_i} V_{h_i-1}^{w_i}(x_i;z_i) \nonumber \\ 
& \quad +  (z-z_i)^{w_i+1} (\gamma_{-w_i-1} V_{h_i}^{w_i})(x_i;z_i) +  \mathcal{O}((z-z_i)^{w_i+2})\ .
\end{align}
We can therefore determine the behaviour of the correlator on the left-hand-side of eq.~(\ref{eq:covering map})
order by order in $z-z_i$ for each $i$, and then compare to the right-hand-side. For example, to the order $\mathcal{O}((z-z_i)^{w_i})$, the left-hand-side of  \eqref{eq:covering map} equals
\begin{align} 
\Bigl\langle \!\gamma(z)\! \prod_{i=1}^n V_{h_i}^{w_i}(x_i;z_i)\! \Bigr \rangle&=x_i \Bigl\langle \prod_{\ell=1}^n V_{h_\ell}^{w_\ell}(x_\ell;z_\ell)\! \Bigr\rangle+(z-z_i)^{w_i}\Bigl\langle V_{h_i-1}^{w_i}(x_i;z_i)\prod_{\ell\ne i}^n V_{h_\ell}^{w_\ell}(x_\ell;z_\ell)\! \Bigr\rangle\nonumber\\
&\qquad+\mathcal{O}((z-z_i)^{w_i+1})  \\
&=(x_i+a_i^{\Gamma}(z-z_i)^{w_i}) \Bigl\langle \prod_{\ell=1}^n V_{h_\ell}^{w_\ell}(x_\ell;z_\ell) \!\Bigr\rangle+\mathcal{O}((z-z_i)^{w_i+1})\ , \!\! 
\end{align}
where we have used the solution \eqref{eq:localisation solution}. This therefore equals the right-hand-side of \eqref{eq:covering map} to this order. We have also checked the equality to the next order. To this end we have used the Wakimoto representation to rewrite 
\begin{align}
\gamma_{-w_i-1} V_{h_i}^{w_i}&=-\frac{1}{k+2j_i}\left(J^+_{w_i-1} V_{h_i-2}^{w_i}-2J^3_{-1} V_{h_i-1}^{w_i}+J^-_{-w_i-1}V_{h_i}^{w_i}\right)\ .
\end{align}
Inserting this into the correlator 
\be \label{eq:gamma subleading term}
\Bigl\langle (\gamma_{-w_i-1}V_{h_i}^{w_i})(x_i;z_i)\prod_{\ell\ne i}^n V_{h_\ell}^{w_\ell}(x_\ell;z_\ell) \Bigr \rangle \ ,
\ee
and using the usual contour techniques, we can rewrite this in terms of a contour integral of correlators with the insertion of $J^a(z)$, which we know how to compute.\footnote{Because of the presence of the secret representations we cannot use the usual contour deformation arguments for the Wakimoto fields themselves, but only for the $\mathfrak{sl}(2,\mathds{R})$ currents.} We have confirmed explicitly for low values of $w_i$ that this then reproduces the corresponding term in the expansion of the actual covering map.  We should stress that in this calculation we did not assume from the outset that the correlator \eqref{eq:covering map} had additional poles, nor made any assumption about what their residues are (except that, if they exist, they are invisible to the $\mathfrak{sl}(2,\mathds{R})$ currents so that we can perform the usual contour deformation arguments for the currents).

\subsection{The correlators of \texorpdfstring{$\partial\Phi$}{dPhi}}

It is similarly very instructive to compute the correlator with the field $\partial\Phi(z)$ inserted. Using essentially the same methods as for the calculation of the correlator with $\gamma(z)$, we have found that the answer is of the form 
\be 
\Bigl\langle \partial\Phi(z) \prod_{i=1}^n V_{h_i}^{w_i}(x_i;z_i) \Bigr\rangle=\left(\sum_{i=1}^n \frac{\frac{j_i}{k}-\frac{1}{2}w_i}{z-z_i}+\sum_{a=1}^N \frac{1}{z-z_a^*}\right) \Bigl\langle \prod_{i=1}^n V_{h_i}^{w_i}(x_i;z_i) \Bigr\rangle\ , \label{eq:dPhi correlator}
\ee
where the $z_1^*, \ldots, z_N^*$ are again the positions of the poles of the covering map $\Gamma(z)$. 
We should mention that it is very natural that the correlator \eqref{eq:dPhi correlator} only has first order poles in $z$ since $\partial \Phi$ is a spin one field. The residue at $\infty$ of this $\partial \Phi(z)$ correlator is then
\begin{align} 
\mathop{\text{Res}}_{z=\infty} \frac{\left\langle \partial\Phi(z) \prod_{i=1}^n V_{h_i}^{w_i}(x_i;z_i) \right \rangle}{\left\langle \prod_{i=1}^n V_{h_i}^{w_i}(x_i;z_i) \right \rangle}&=\sum_{i=1}^n \left(-\frac{j_i}{k}+\frac{1}{2}w_i\right)-N\\
&=-\sum_{i=1}^n \frac{j_i}{k}-\frac{1}{2}(2-n)=-\frac{1}{k}\ ,
\end{align}
where we have used the Riemann-Hurwitz formula \eqref{eq:degree covering map}, as well as the condition on the $j$'s of eq.~\eqref{eq:j condition}. This is what one should have expected since $\Phi$ has a background charge, which leads to an anomalous conformal transformation behaviour (see Appendix~\ref{app:anomalous partialPhi}), and in turn to the charge conservation law 
\be 
\sum_i Q_i=-\frac{1}{k}\ .\label{eq:anomalous charge conservation sphere}
\ee 
Thus we can think of the condition on the $j$'s in eq.~\eqref{eq:j condition} as the charge conservation condition for $\partial \Phi$. If it were not satisfied, one has to include screening charges in the correlator to compute them in the Wakimoto representation. This would make the correlator non-holomorphic in $z$ and thus $\gamma(z)$ would lose its interpretation as a covering map. 

Finally, we note that the residue of $\partial \Phi$ at $z_\ell^*$ is always $1$. We also give some explanation of this in Appendix~\ref{app:anomalous partialPhi}.

\subsection{The ground state solution}

We recall from Section~\ref{sec:jcond} that a natural solution to eq.~\eqref{eq:j condition} is given by setting, see (\ref{jsol})
\be 
j_1=1-\frac{k}{2}\ , \quad \qquad  j_i=\frac{k}{2} \ ,\quad i\geq 2 \ ,
\ee 
and that for these values of $j_i$ the vertex operators describe the ground states of the twisted sector representations of the dual symmetric orbifold CFT \cite{Dei:2019osr}. 
If we place the conjugate field (i.e.\ the one for which $j_1=1-\frac{k}{2}$) at $z=\infty$, the $\partial \Phi(z)$ correlator actually simplifies to 
\begin{align} 
\Bigl\langle \partial\Phi(z) \prod_{i=1}^n V_{h_i}^{w_i}(x_i;z_i) \Bigr\rangle&=-\frac{1}{2}\left(\sum_{i=2}^n \frac{w_i-1}{z-z_i}-\sum_{a=1}^N \frac{2}{z-z_a^*}\right) \Bigl\langle \prod_{i=1}^n V_{h_i}^{w_i}(x_i;z_i) \Bigr\rangle \\
&=- \frac{\partial^2 \Gamma(z)}{2\,\partial \Gamma(z)}\Bigl\langle \prod_{i=1}^n V_{h_i}^{w_i}(x_i;z_i) \Bigr\rangle\ .\label{eq:dPhi ground states}
\end{align}
Thus, we can also express the $\partial \Phi(z)$ correlator in a very simple way in terms of the covering map. This formula has an important interpretation, see Sections~\ref{sec:classquant} and \ref{sec:onshell}.

\subsection{Relation to the Schwarzian}

Let us end this section by noting that the correlators of the currents are actually closely related to the Schwarzian of the covering map. In order to see this, we take the conformal weights to match the value of the twisted sector ground states of the symmetric product orbifold
\be 
h_i=\frac{k(w_i^2-1)}{4w_i}\ ,
\ee
and choose the spins as in \eqref{jsol}. Then we observe experimentally that
\begin{subequations}
\begin{align}\label{Jpclass}
\left\langle J^+(z) \prod_{i=1}^n V_{h_i}^{w_i}(x_i,z_i) \right\rangle & =-\frac{k}{2} \frac{S[\Gamma](z)}{\partial \Gamma(z)}\left\langle  \prod_{i=1}^n V_{h_i}^{w_i}(x_i,z_i) \right\rangle + \partial_z \Bigl( \cdots \Bigr) \\[4pt]
\left\langle J^3(z) \prod_{i=1}^n V_{h_i}^{w_i}(x_i,z_i) \right\rangle & =-\frac{k}{2} \frac{\Gamma(z)\, S[\Gamma](z)}{\partial \Gamma(z)}\left\langle  \prod_{i=1}^n V_{h_i}^{w_i}(x_i,z_i) \right\rangle + \partial_z \Bigl( \cdots \Bigr) \\[4pt]
\left\langle J^-(z) \prod_{i=1}^n V_{h_i}^{w_i}(x_i,z_i) \right\rangle & =-\frac{k}{2} \frac{\Gamma(z)^2\,  S[\Gamma](z)}{\partial \Gamma(z)}\left\langle  \prod_{i=1}^n V_{h_i}^{w_i}(x_i,z_i) \right\rangle + \partial_z \Bigl( \cdots \Bigr) \ ,
\end{align}
\end{subequations}
where the functions in $\bigl( \cdots \bigr)$ are single-valued, and $S[f]$ denotes the Schwarzian,
\be
S[f](z) = \frac{f'''(z)}{f'(z)} - \frac{3}{2} \Bigl(  \frac{f''(z)}{f'(z)} \Bigr)^2 \ . 
\ee
Interpreted in terms of classical fields, this means for example
\be
J^+(z) = -\frac{k}{2} \frac{S[\Gamma](z)}{\partial \Gamma(z)} + \partial_z \Bigl( \cdots \Bigr) \ , 
\ee
and similarly for the other cases. Hence, we have, in particular, that
\begin{subequations}
\begin{align}
J^+_0&=\oint_0 \mathrm{d}z \ J^+(z) =-\frac{k}{2} \oint_0 \mathrm{d}z\ \frac{S[\Gamma](z)}{\partial \Gamma(z)}\ , \\
J^3_0&=\oint_0 \mathrm{d}z \ J^3(z) =-\frac{k}{2} \oint_0 \mathrm{d}z\ \frac{S[\Gamma](z)}{\partial \Gamma(z)}\, \Gamma(z)\ , \\
J^-_0&=\oint_0 \mathrm{d}z \ J^-(z) =-\frac{k}{2} \oint_0 \mathrm{d}z\ \frac{S[\Gamma](z)}{\partial \Gamma(z)}\, \Gamma(z)^2 \ .
\end{align}
\end{subequations}
These generators are to be identified with the M\"obius generators $L_{-1}$, $L_0$ and $L_1$ of the dual CFT.\footnote{Note that in (\ref{Jpclass}) the $J^a(z)$ refer to the decoupled bosonic currents. However, since the $V_{h_i}^{w_i}(x_i,z_i)$ correspond to spectrally flowed affine primary states, the result would be the same if we replaced them by the full superaffine currents that are relevant for the definition of the dual CFT M\"obius generators.}

In fact, one can construct the entire `dual' Virasoro algebra of Brown \& Henneaux \cite{Brown:1986nw} within the $\mathrm{SL}(2,\mathds{R})$ 
WZW model by defining \cite{Giveon:1998ns}
\be \label{Virspace}
\mathcal{L}_m=k \oint \mathrm{d}z \left(-(m+1)\gamma^m \partial\Phi+\gamma^{m+1} \beta\right)(z) = -\frac{k}{2}\oint \mathrm{d}z \ \frac{S[\Gamma](z)}{\partial \Gamma(z)}\, \Gamma(z)^{m+1} \ ,
\ee
where the last identity holds for this classical solution. Eq.~(\ref{Virspace}) now has a nice interpretation: since the energy-momentum tensor of a CFT is a quasiprimary field, it has the anomalous transformation property that 
\be\label{Virtrans}
T\bigl(f(z)\bigr) = \bigl(\partial f(z)\bigr)^{-2}\,\left( T(z) - \frac{c}{12} \, S[f](z) \right) \ , 
\ee
where $f(z)$ is any analytic function. Thus we can write (\ref{Virspace}) as 
\be
\mathcal{L}_m = \int \mathrm{d}x \ T^{\rm st}(x) x^{m+1} = \int \mathrm{d}z\  \Bigl( T^{\rm cov}(z) - \frac{c}{12} S[\Gamma](z) \Bigr) \bigl( \partial \Gamma(z) \bigr)^{-1} \, \Gamma(z)^{m+1}\ , 
\ee
where $T^{\rm st}$ is the spacetime stress-energy tensor, $T^{\rm cov}$ its lift to the covering surface, and they are related to one another via (\ref{Virtrans}) with $x = \Gamma(z)$. We have also used that for the ground states of the twisted sectors $T^{\rm cov}=0$, together with $\frac{c}{12} = \frac{k}{2}$ since $c=6k$. Thus the full spacetime Virasoro algebra arises essentially just from the Schwarzian transformation behaviour of the energy-momentum tensor.

\section{The classical solution}\label{sec:classical}

In this section we will start again by studying classical solutions of string theory on $\text{AdS}_3$. We are, in particular, interested in the solutions which correspond to the ground states in the $w$-spectrally flowed sector with $j=\frac{1}{2}$. They sit at the bottom of what is usually a long string continuum, but are actually the only allowed representations for $k=1$ \cite{Eberhardt:2018ouy}. As we will see, they can be obtained by taking a suitable limit of a family of solutions that localises them to the boundary of the $\text{AdS}_3$ space --- this is natural since they have no radial momentum. This is also the reason, as we will see, why the theory at $k=1$ can be studied semi-classically on the worldsheet despite describing a highly curved spacetime. 
The limiting solutions will have the right behaviour near the boundary, and we show in 
Section~\ref{sec:classquant} that the same solutions emerge also from the quantum analysis of correlators that was performed in Sections~\ref{sec:simple solution} and \ref{sec:covering}. Furthermore, we explain in Section~\ref{sec:onshell} that their worldsheet action (\ref{adsact}) reproduces exactly the Liouville action of  Lunin-Mathur  (\ref{liouv}), as already sketched in Section~\ref{sec:Goal}.

\subsection{Classical Solutions at the Boundary of \texorpdfstring{$\text{AdS}_3$}{AdS3}} \label{subsec:classical sol}

The bosonic action on $\text{AdS}_3$ given in (\ref{adsact}) 
can also be written in a first order form as
\be\label{adsact2}
S_{\mathrm{AdS}_3}= \frac{k}{4\pi} \int \mathrm{d}^2 z\  (4 \partial \Phi \bar{\partial} \Phi  +\bar \beta\partial  \bar{\gamma} +\beta \bar{\partial}  \gamma - \mathrm{e}^{-2\Phi} \beta \bar\beta-k^{-1} R \Phi)  \ . 
\ee
Here the equations of motion fix $\beta =\mathrm{e}^{2\Phi}\partial  \bar{\gamma}$ and $\bar\beta =\mathrm{e}^{2\Phi} \bar\partial  \gamma$, and hence reduce (\ref{adsact2}) to (\ref{adsact}). The last term arises from a quantum renormalisation. In terms of these fields, the group elements can be written as --- we are using the conventions of \cite{Ponsot:2001gt}
\be\label{gdef}
g = \left( \begin{matrix} e^{\Phi} &  e^{\Phi}\, \bar{\gamma} \\ 
 e^{\Phi} \, \gamma & e^{\Phi} \gamma \bar{\gamma} + e^{-\Phi} \end{matrix} \right) \quad \hbox{with} \quad 
g^{-1}  = \left( \begin{matrix} e^{\Phi} \gamma \bar{\gamma} + e^{-\Phi}  & 
- e^{\Phi} \, \bar{\gamma} \\  - e^{\Phi} \, \gamma & e^{\Phi} \end{matrix} \right)  \ , 
\ee
and the currents are defined via 
\be
- k\, \partial g \, g^{-1} =  \left( \begin{matrix}  J^3 &  - J^+ \\   J^- & - J^3 \end{matrix} \right) \qquad \hbox{and} \qquad 
k\, g^{-1} \bar\partial g   =  \left( \begin{matrix}  \bar{J}^3 &  - \bar{J}^+ \\   \bar{J}^- & - \bar{J}^3 \end{matrix} \right) \ .
\ee 
Thus  these variables can be identified with the Wakimoto fields given in eq.~(\ref{eq:Wakimoto}).\footnote{The shift $k \to k+2$ in the formula for $J^-$ is a result of a correction in the quantum theory due to normal ordering.} The equations of motion then take the form 
\begin{subequations}
\begin{align}
\bar\partial \beta & =   0 \\ 
-\bar\partial \partial \Phi + \beta \, \bar\partial \gamma & =  0 \label{2nd} \\
2\, \partial\Phi \bar\partial\gamma + \bar\partial\partial\gamma & =  0 \ , \label{3rd}
\end{align}
\end{subequations}
and similarly for $\bar{\beta}$ and $\bar{\gamma}$. 
The general solution to these equations of motion can be parametrised by three holomorphic ($\rho(z)$, $b(z)$ and $a(z)$) and three anti-holomorphic ($\bar\rho(\bar{z})$, $\bar{b}(\bar{z})$ and $\bar{a}(\bar{z})$) functions. In the notation of \cite{deBoer:1998gyt}, we have 
\begin{subequations}\label{eq:classsol}
\begin{align}
\Phi(z,\bar{z}) &= \rho(z) +\bar{\rho}(\bar{z})+\log\big(1+b(z)\bar{b}(\bar{z})\big) \ ,\\
\gamma(z,\bar{z}) &= a(z) + \frac{\mathrm{e}^{-2\rho(z)}\bar{b}(\bar{z})}{1+b(z)\bar{b}(\bar{z})} \ ,\\
\bar{\gamma} (z,\bar{z})&= \bar{a}(\bar{z}) + \frac{\mathrm{e}^{-2\bar{\rho}(\bar{z})}b(z)}{1+b(z)\bar{b}(\bar{z})}\ ,
\end{align}
\end{subequations}
where 
\be
\beta(z) = \mathrm{e}^{2\rho(z)} \, \partial b (z) \ , \qquad \bar\beta (\bar{z}) = \mathrm{e}^{2\bar\rho(\bar{z})} \, \bar\partial \bar{b} (\bar{z})\ . 
\ee
Thus the fields $\gamma$ and $\bar{\gamma}$ are not, in general, holomorphic or anti-holomorphic, respectively, and $\Phi$ is, in general, not a sum of a holomorphic and an anti-holomorphic function. 

However, there is a limit one can take in which we scale 
\be\label{scalim}
b(z) = b_0 + \epsilon \, c(z) \ ,\qquad \rho(z) = - \frac{1}{2}  \log\epsilon + \sigma(z) \ , 
\ee
and similarly for $\bar{b}(\bar{z})$ and $\bar{\rho}(\bar{z})$, and then take $\epsilon \rightarrow 0$, while keeping 
$b_0$ constant, and $c(z), \sigma(z)$ finite. In this limit we find 
\begin{subequations}\label{limsol}
 \begin{align}
\Phi(z,\bar{z}) &= -\log{\epsilon} + \log(1 + b_0 \bar{b}_0) +  \sigma(z) +\bar{\sigma}(\bar{z}) \\
\beta(z) &= \mathrm{e}^{2\sigma(z)} \, \partial c (z)  \\
\gamma(z) &=a(z)  \ . 
\end{align}
\end{subequations}
Thus $\beta$ and $\gamma$ are now holomorphic (and similarly $\bar\beta$ and  $\bar\gamma$ are anti-holo\-mor\-phic), and $\Phi$ is a sum of a holomorphic and an anti-holomorphic function. This solution reflects the fact that $\Phi$ has an infinite additive constant ($-\log{\epsilon}$) which essentially places the worldsheet at the boundary. As can be seen from (\ref{adsact2}) the term $\beta\bar{\beta} \mathrm{e}^{-2\Phi}$ drops out in this limit and we are left with a free theory \cite{deBoer:1998gyt} (which is, in particular, semiclassically exact). 

In order to understand this limit further, let us look at a family of solutions that describe the two-point functions $\langle V_h^w(\infty;\infty) \, V_h^w(0;0) \rangle$ for spectrally flowed affine primary states in the $w$-spectrally flowed sector. They are characterised by the property that the currents only have their (spectrally flowed) zero modes excited and therefore take the form 
\be\label{eq:Jansatz}
J^3(z) = \frac{h}{z}\ , \qquad J^\pm (z) = \alpha \, z^{\mp w-1}\ ,
\ee
where $h$ and $\alpha$ are constants. We will find it convenient to parametrise them as 
\be
\alpha=\frac{kp}{\nu}\ , \qquad h =\frac{k}{2}\left(p\, \frac{\sqrt{\nu^2+4}}{\nu}+ w\right)\ .
\ee
The mass-shell condition requires that $h^2=\alpha^2$, which allows us to solve for $\nu$ in terms of $p$ as 
\be
\nu = - \frac{4wp}{w^2-p^2} \ ,
\ee
provided that $w^2\neq p^2$. Up to one integration constant, which we have set to zero, the most general solution of the form (\ref{eq:Jansatz}), fixes the functions entering into (\ref{eq:classsol}) as 
\begin{subequations}\label{eq:abrho}
\begin{align}
a(z)&= \frac{1}{2} z^w \left(\sqrt{\nu^2+4}+\nu\right)\ , \\
\rho (z)&=  - \frac{1}{2} \log{\epsilon}+\frac{1}{2} (p-w) \log (z)\ , \\
b(z)&= b_0 -  \frac{\epsilon}{\nu}\, z^{-p}\ .
\end{align}
\end{subequations}
Here we have suggestively denoted the two remaining constants in the solution by $\epsilon$ and $b_0$, which puts them precisely in the form of eq.~(\ref{scalim}). 
 
We can now implement the scaling limit of $\epsilon\rightarrow 0$ on this family of solutions --- as discussed, this is the limit where the solutions have support only on the boundary. In this limit, the
expressions for the Wakimoto fields $\gamma$ and $\Phi$ become holomorphic 
\begin{subequations}\label{Phisol}
\begin{align}
\gamma(z)&=\frac{1}{2}\left(\nu+\sqrt{\nu^2+4}\right)z^w\ , \\
\Phi(z,\bar{z})&= -\log\epsilon - \tfrac{(w-p)}{2}\log(z) - \tfrac{(w-p)}{2} \log(\bar{z})\ . 
\end{align}
\end{subequations}
By comparison with (\ref{eq:dPhi correlator}) we see that the parameter $p$ should be identified with the spin of the affine primary as\footnote{Note that this identification does not need our results obtained from the quantum analysis. It is simply a consequence of the Wakimoto representation of affine primaries, see e.g.~\cite[eq.~(2.12)]{Eberhardt:2019qcl}.} 
\be
p = \frac{2j}{k} \ . 
\ee
For $k=1$ and $j=\frac{1}{2}$, we should thus take $p=1$, and then $\nu$ and $h$ equal
\be
\nu = - \frac{4w}{w^2-1} \ , \qquad h = \frac{k (w^2-1)}{4w} \ . 
\ee
Note that $h$ is indeed the ground state energy of the $w$-twisted sector, as expected. For this choice of $p$, $\gamma(z)$ and $\Phi(z)$ then simplify to
\be\label{eq:finalsol}
\gamma(z) = \frac{w-1}{w+1} \, z^w  \ , \qquad 
\Phi(z,\bar{z})= -\log\epsilon - \tfrac{(w-1)}{2}\log(z) - \tfrac{(w-1)}{2} \log(\bar{z}) \ ,  
\ee
thus reproducing eq.~(\ref{coversol}). (The prefactor in the definition of $\gamma(z)$ is immaterial, since for a two-point function the little-group M\"obius symmetry is non-trivial, which allows one to change the scaling of the covering map.)

We should mention in passing that the class of solutions found by Maldacena and Ooguri \cite{Maldacena:2000hw} --- the so-called ``spacelike" solutions of their Section~2.4  --- do not quite fit into the above description. While they also lead to (\ref{eq:Jansatz}), they cannot be written in the form of eq.~(\ref{eq:abrho}) since the relevant integration constant is not zero for them. In any case, they do not describe the ground state of the twisted sector since their conformal dimension is always larger than $h\geq \frac{kw}{4}$, see eq.~(41) of \cite{Maldacena:2000hw}. This can also seen geometrically since their solution corresponds to the pulsating profile of Figure~4 of \cite{Maldacena:2000hw}, whereas the ground state should not have nodes, but rather be a simple cylinder (at least for $w=1$).

\subsection{Relation to quantum correlators}\label{sec:classquant}

We argued below eq.~(\ref{limsol}) that for solutions  that are at the boundary, we have a semiclassically exact description. In the previous subsection we have found the simplest case of such a classical configuration which describes a two point function, see  eq.~(\ref{eq:finalsol}).
Thus for a correlator such as (\ref{gencorr}), we must have a classical solution which behaves like (\ref{eq:finalsol}) in the immediate vicinity of each of the insertions $z_i$. Since $\gamma$ is furthermore holomorphic at the boundary, it follows that $\gamma$ must actually agree with the covering map.

Since the solutions of the quantum theory we have studied in Sections~\ref{sec:simple solution} and \ref{sec:covering} are localised at the boundary, they should therefore directly have a classical interpretation. In particular, the quantum correlators (\ref{eq:covering map}) and (\ref{eq:dPhi ground states}) lead to the identifications  
\begin{align}\label{classcov}
\gamma(z)&=\Gamma(z)\ , \\
\partial \Phi(z)&= -\frac{\partial^2 \Gamma(z)}{2\,\partial \Gamma(z)}\ , \label{eq:LMformula}
\end{align} 
where $\Gamma(z)$ is, as before, the relevant covering map. (The expression for $k \beta = J^+$ is more complicated.) 
This is consistent with the solution discussed in Section~\ref{subsec:classical sol}, which corresponds to the covering map $\Gamma(z)=z^w$. Furthermore, since near each $z_i$, the covering map behaves as $\Gamma(z) \sim (z-z_i)^{w_i}$, we have
\be
\partial \Phi(z) = -\frac{\partial^2 \Gamma(z)}{2\,\partial \Gamma(z)} \sim - \frac{(w_i-1)}{2} \, \frac{1}{(z-z_i)} \ , 
\ee
which therefore agrees precisely with eq.~(\ref{eq:finalsol}). Thus the correlation functions we have found in the quantum analysis lead indeed to the correct classical solutions!
\smallskip

We can also express the corresponding group valued fields in terms of the Wakimoto representation, see eq.~(\ref{gdef}) 
\be
g = \left( \begin{matrix} e^{\Phi} \quad &  e^{\Phi}\, \bar{\Gamma} \\ 
 e^{\Phi} \, \Gamma \quad & e^{\Phi} \Gamma \bar{\Gamma} + e^{-\Phi} \end{matrix} \right) 
= \begin{pmatrix}
\frac{1}{\sqrt{\partial \Gamma}} & 0 \\
\frac{\Gamma}{\sqrt{\partial \Gamma}} & \sqrt{\partial \Gamma}
\end{pmatrix}\begin{pmatrix}
\frac{1}{\sqrt{\bar{\partial} \bar\Gamma}} & \frac{\bar\Gamma}{\sqrt{\bar\partial \bar\Gamma}} \\
0 & \sqrt{\bar\partial \bar\Gamma}
\end{pmatrix}\ ,
\ee
where $\bar{\Gamma}(\bar{z})$ is the covering map for the anti-holomorphic variables, and we note that 
\be
\Phi = -\frac{1}{2} \log (\partial\Gamma) - \frac{1}{2} \log (\bar\partial\bar\Gamma) +\text{constant} \ , 
\ee
as follows from (\ref{eq:LMformula}) (together with its anti-holomorphic analogue). As we have discussed in Subsection~\ref{subsec:classical sol}, we expect this constant to be infinite, since the worldsheet is pinned to the boundary.

\subsection{The on-shell action}\label{sec:onshell}

Now that we have assembled all the pieces, we can finally make a quantitative comparison between the worldsheet and the classical action. Recall that the quantum correlators correspond to solutions of the type analysed in eqs.~(\ref{limsol}) that are localised at the boundary of ${\rm AdS}_3$. Now the remarkable property of our solution (\ref{classcov}) is that 
\be\label{solscale} 
\mathrm{e}^{-2{\Phi}}= C \ \big| \partial_z \Gamma\big|^2 \ ,
\ee
where $C$ is a constant, 
since eq.~(\ref{eq:LMformula}) implies that 
\be
\Phi(z,\bar{z}) = {\rm const} - \frac{1}{2} \log (\partial \Gamma) - \frac{1}{2} \log (\bar{\partial} \bar{\Gamma}) \ . 
\ee
In particular, comparing with the conformal factor in the Lunin-Mathur approach, see eq.~(\ref{covscale}), we deduce that the Liouville field $\phi$ of Lunin-Mathur is to be identified with the Wakimoto field $\Phi$ as 
\be
\partial\phi(z)=-2\partial\Phi(z)\ , 
\ee
see eq.~(\ref{scalematch}). Furthermore, as was also already mentioned there, this identification (together with that of the orbifold CFT covering space with the worldsheet) leads to the equality of the semiclassically exact worldsheet sigma model action (\ref{adsact}) (or rather its first order form, see eq.~(\ref{adsact2})), with the Lunin-Mathur Liouville type action (\ref{liouv}) which computes the orbifold correlator. This makes therefore manifest that the two calculations agree.

\section{The generalisation to higher genus}\label{sec:genus}

The above analysis suggests naturally a generalisation to higher genus which it would be interesting to work out in detail. We have not yet tried to do so rigorously, but many properties we have investigated seem to have a direct generalisation. In particular, the semiclassical argument of the previous section does not depend on the topology of the worldsheet and seems to go through. Let us therefore assume that the worldsheet theory is defined on a higher genus Riemann surface $\Sigma_g$, though we keep the boundary space (parametrised by $x$) to still be a sphere.\footnote{In \cite{Eberhardt:2018ouy} we considered the one-loop worldsheet CFT partition function where the spacetime was thermal ${\rm AdS}_3$ and hence the boundary was a torus. We found in that case a delta-function localisation to points in the moduli space which admit holomorphic covering maps, now from the worldsheet $\mathbb{T}^2$ to the spacetime boundary $\mathbb{T}^2$, see the discussion around eq.~(4.27) of \cite{Eberhardt:2018ouy}.}

\subsection{The constraint on the spins}
Let us start by discussing the constraint on the spins $j_i$ that we expect in general. The anomalous charge conservation law of the Wakimoto field $\partial \Phi(z)$ has an immediate generalisation to higher genus surfaces. The anomalous charge comes from the term 
\be 
-\frac{1}{4\pi}\int \mathrm{d}^2 z \, \sqrt{g}\,R \Phi
\ee
in the worldsheet action \eqref{adsact2}, which, by the Gauss-Bonnet theorem, implies that the anomalous conservation  law is proportional to the Euler characteristic $\chi(\Sigma_g)$ of the surface. Thus eq.~\eqref{eq:anomalous charge conservation sphere} becomes in general, see e.g.~\cite[Section 9.1.3]{DiFrancesco:1997nk}
\be 
\sum_i Q_i =-\frac{1}{2k} \, \chi (\Sigma_g)=\frac{g-1}{k}\ .
\ee
On the other hand, the residue of the field $\partial \Phi(z)$ at the insertion points should equal $\frac{j_i}{k}-\tfrac{1}{2}w_i$, while that at the poles of $\Gamma(z)$ should be $1$; both are local properties and hence should not be affected by the genus. Finally, we recall that the Riemann-Hurwitz formula becomes for generic genus $g$ 
\be 
N=\frac{1}{2} \sum_{i=1}^n (w_i-1)+1-g  = \frac{1}{2} \sum_{i=1}^{n} w_i - \frac{1}{2} (n-2+2g)\ . \label{eq:Riemann Hurwitz formula}
\ee
Then requiring the anomalous charge conservation law, i.e. 
\be
\sum_{i=1}^n \left( \frac{j_i}{k} - \frac{w_i}{2} \right) + N = \sum_{i=1}^n \frac{j_i}{k} - \frac{1}{2} (n-2+2g) = \frac{1-g}{k} \ , 
\ee 
leads to 
\be 
\sum_{i=1}^n j_i=\frac{(k+2)}{2}(n-2+2g)-\text{dim}(\mathcal{M}_{g,n})\ , \label{eq:general j constraint}
\ee
where $\text{dim}(\mathcal{M}_{g,n})=3g-3+n$ is the dimension of the moduli space of genus $g$ Riemann surfaces with $n$ punctures. Note that $\hat{k}=(k+2)$ is the level of the bosonic $\mathfrak{sl}(2,\mathds{R})$ algebra, and eq.~(\ref{eq:general j constraint}) is therefore the generalisation of eq.~(\ref{eq:j condition}) to the general case. As was explained there, the second term arises from the presence of $\delta$-functions in the correlation function, thus suggesting that in general we will have $\text{dim}(\mathcal{M}_{g,n})$ many $\delta$-functions, i.e.\ that the integral over moduli space will again be completely localised.

It is also striking that this constraint is always satisfied for the situation of primary interest, where $j_i=\tfrac{1}{2}$ and $k=1$,  see also the discussion in Section~\ref{sec:jcond}.

\subsection{The correlation functions}
It is therefore again natural to assume that the correlation functions localise, i.e.~take the form
\be \label{eq:highergenuscov}
\left\langle \prod_{i=1}^n V_{h_i}^{w_i}(x_i,z_i)  \right\rangle= \sum_\Gamma\delta^{(3g-3+n)}(f_\Gamma(x,z,\tau)) \prod_{i=1}^n (a_i^\Gamma)^{-h_i} W_\Gamma(x_1,\dots, x_n)\ ,
\ee
where $f_\Gamma=0$ defines the appropriate localisation constraint so that the worldsheet coordinates and worldsheet moduli localise on all possible branched covering surfaces. In this context it is actually more convenient to express the remaining freedom in the function $W_\Gamma$ as a function of $(x_1,\dots,x_n)$, since this is the correct number of variables that remain after the $\delta$ function constraints have been imposed. Finally, this number can be cut down further  to $n-3$, using the global $\mathrm{SL}(2,\mathds{R})$ Ward identities.

As an example, let us consider the torus covering the sphere branched over four points with ramification index 2. The covering map is explicitly given by
\be 
\Gamma(z)=\frac{x_1(x_2-x_3)\wp(z;\tau)+x_3(x_1-x_2)\wp(\frac{1}{2};\tau)+x_2(x_3-x_1)\wp (\frac{\tau}{2};\tau)}{(x_2-x_3)\wp(z;\tau)+(x_1-x_2)\wp(\frac{1}{2};\tau)+(x_3-x_1)\wp (\frac{\tau}{2};\tau)}\ , \label{eq:covering map torus}
\ee
where $\wp(z;\tau)$ is the Weierstrass $\wp$ function. The four ramification points of order two are  given by $z_1=0$, $z_2=\tfrac{1}{2}$, $z_3=\tfrac{\tau}{2}$ and $z_4=\tfrac{\tau+1}{2}$. They are mapped to $x_1$, $x_2$, $x_3$ and $x_4$ respectively, where
\be 
x_4=\Gamma(\tfrac{\tau+1}{2})=\frac{-x_2 x_3 \vartheta_3(\tau)^4+x_1x_3 \vartheta_4(\tau)^4+x_1x_2 \vartheta_2(\tau)^4}{x_1 \vartheta_3(\tau)^4-x_2 \vartheta_4(\tau)^4-x_3 \vartheta_2(\tau)^4}\ ,
\ee
and the $\vartheta_i(\tau)$ are the usual Jacobi $\vartheta$-functions. Note that the points $z_1, \dots,z_4$ are fixed up to an overall shift in terms of the modulus $\tau$, which in turn is fixed in terms of the cross-ratio $x$. This fixes all moduli of the four-punctured torus and corresponds to the constraint $f_\Gamma(x,z,\tau)$ in this case. We also mention in passing that the covering map \eqref{eq:covering map torus} has two poles, since the Weierstrass $\wp$ function attains every value twice.

\subsection{The classical solution}

Finally, there is also a fairly natural idea for how the classical solutions generalise, at least for the torus case $g=1$. Concentrating again on the ground state solutions, we notice that taking $(n-1+g)$ fields to have spin $\tfrac{k}{2}$ and $1-g$ fields to have spin $1-\tfrac{k}{2}$ solves the general $j$ constraint \eqref{eq:general j constraint}. Note that there is an obvious problem with this if $g\ge 2$, since the second number becomes negative; we are not exactly sure how to cure this (although this should only be an artefact of our parametrisation since the representations $j$ and $1-j$ should be equivalent). This problem disappears in the situation of primary interest, where $k=1$ and $j=\tfrac{1}{2}$.

For instance, in the genus one example from the previous subsection, we still simply have
\be 
\partial \Phi(z)=- \frac{\partial^2 \Gamma(z)}{2\,\partial \Gamma(z)}\ ,
\ee
which has all the required properties. It is maybe worth mentioning that geometrically the situation on the torus is actually easier than on the sphere, since the coordinate $z$ is a globally flat coordinate. This is no longer possible on genus $g \ge 2$ surfaces, which is related to the above problem. It would be interesting to work this out in more detail.

\subsection{String perturbation theory}

The proposed localisation of these higher genus amplitudes to the solutions of the covering map, see eq.~(\ref{eq:highergenuscov}), has a striking consequence: it demonstrates that the worldsheet theory reproduces at least structurally all $1/N$ corrections of the dual CFT. In order to understand this, we recall from \cite{Pakman:2009zz} that the large $N$ scaling behaviour of the symmetric orbifold correlators has a very suggestive form: the contribution of an $n$-point function that comes from a covering surface at genus $g$ goes (in the large $N$ limit) as 
\be\label{symorbcor}
\langle \mathcal{O}_1 \cdots \mathcal{O}_n \rangle \sim N^{1-g-\frac{n}{2}}  \ . 
\ee
Since we should identify the string coupling constant $g_\text{s}$ (again in the large $N$ limit) with 
\be\label{gs}
g_\text{s} \sim \frac{1}{\sqrt{N}} \ ,
\ee
this translates into 
\be
\langle \mathcal{O}_1 \cdots \mathcal{O}_n \rangle \sim g_\text{s}^{2g-2 + n }  \ ,
\ee
and thus suggests that the genus $g$ of the covering map should be identified with the genus of the corresponding worldsheet contribution \cite{Pakman:2009zz}, see also \cite{Lunin:2000yv}. Actually, as stressed in \cite{Pakman:2009zz}, identifying the genus in this manner is the natural correspondence, since there are subleading $1/N$ corrections in eq.~(\ref{symorbcor}), and thus it is difficult to express the correspondence directly in terms of $N$, i.e.\ as in eq.~(\ref{gs}).

Obviously, the simplest way of realising this idea --- this was also already noted in \cite{Lunin:2000yv,Pakman:2009zz} --- would be if the {\it 
worldsheet is the covering surface}, and this is what our analysis has shown! It establishes that this beautiful picture is correct, and in effect, guarantees that all higher genus (or $1/N$) corrections are correctly reproduced by the worldsheet theory. Furthermore, it leads to the following non-re\-nor\-ma\-li\-sation theorem:
\begin{verse} 
\hspace{0.5cm} \textit{Any $n$-point function of spectrally flowed affine primaries of the \linebreak $\mathrm{SL}(2,\mathds{R})$ WZW model vanishes identically on a Riemann surface of sufficiently high genus. Moreover, two-point functions are only non-vanishing on the sphere. In particular, the string theoretic two-point functions are independent of the string coupling constant $g_\mathrm{s}$.}
\end{verse}

\noindent 
This is easily proven from the Riemann-Hurwitz formula \eqref{eq:Riemann Hurwitz formula}
\begin{align}\label{RHp}
2-2g=2N-\sum_{i=1}^n (w_i-1) \ge 2 \ \text{max}(w_i)-\sum_{i=1}^n(w_i-1)\ ,
\end{align}
by noting that $N\geq \text{max}(w_i)$. Hence a covering map can only exist for\footnote{A more refined argument shows that this is the optimal bound; it follows from basic Hurwitz theory that a covering map with $N=\text{max}(w_i)$ always exists.}
\be 
g \le 1-\text{max}(w_i)+\sum_{i=1}^n \frac{w_i-1}{2}\ ,\label{eq:genus inequality}
\ee
and thus the correlation function can only be non-vanishing for finitely many genera. For $n=2$, we have necessarily $N=w_1=w_2$ and hence $g=0$. In particular, this explains why the spectrum of the dual CFT that was matched exactly from a tree-level worldsheet analysis in \cite{Eberhardt:2018ouy} does not get any higher genus corrections.\footnote{The spectrum of the symmetric product orbifold stabilises in the large $N$ limit. Hence the symmetric product orbifold at finite $N$ has \emph{less} states than the limit $N \to \infty$. The disappearance of states is non-perturbative in the large $N$ expansion and is not captured by our worldsheet analysis.}

\section{Conclusions}\label{sec:concl}

What is fascinating about the particular example of the $\text{AdS}/\text{CFT}$ correspondence we have studied in this paper, is its potential to demystify many aspects of gauge-string duality. Here is a case where one can look into the machine, so to say, and see how the nuts and bolts fit together and how the gears move. We have tried to lay bare the mechanism which underlies the equivalence of a well defined worldsheet string theory on $\text{AdS}$ to a spacetime CFT on the boundary. An advantage is that we did not need to explicitly compute observables on either side to see the equivalence. 

It helped, of course, that we were working in the ``tensionless" limit of a small radius $\text{AdS}$. The theory essentially becomes a topological string theory in the $\text{AdS}$ (as well as $\text{S}^3$ directions, as was observed in \cite{Eberhardt:2018ouy}) as it is too ``small" to support physical transverse excitations. But there are genuine physical string oscillators on the $\mathbb{T}^4$. It thus exhibits a complexity intermediate between purely topological open-closed string dualties such as those of \cite{Gopakumar:1998ki, Gaiotto:2003yb, Razamat:2008zr, Gopakumar:2011ev} and those with propagating gravity.\footnote{Note, however, that the dual symmetric product CFTs are complex enough to capture the microstates and other physics of 5d black holes \cite{Strominger:1996sh, Maldacena:1996ky, David:2002wn}. It would be very interesting to see and understand this (non-perturbative) sector explicitly from the worldsheet point of view. Here we have been working strictly in a genus expansion.}

Many of the features that we see in the present example are what one might also expect of the dual to free gauge theories in other dimensions \cite{Sundborg:2000wp, Witten, Mikhailov:2002bp}. In the general programme of reconstructing worldsheet theories dual to free field theories \cite{Gopakumar:2003ns, Gopakumar:2004qb, Gopakumar:2005fx}, a similar truncation of the genus expansion was found \cite{Aharony:2006th}, as were delta function and other distributional features on moduli space \cite{Aharony:2007fs, Razamat:2008zr, Gopakumar:2011ev, Gopakumar:2012ny}; both of these are hallmarks of topological string theories \cite{Distler:1989ax, Verlinde:1990ku, Bershadsky:1993ta}. This may be an opportune time to revisit these considerations and renew the push to connect with topological descriptions of tensionless $\text{AdS}$ worldsheet theories with RR-flux \cite{Berkovits:2008qc, Berkovits:2019ulm}. Perhaps one might even be able to embed the present duality in higher dimensions and capture a sector thereof as in \cite{Costello:2018zrm}. We cannot also resist pointing out the close resemblance to the dual string description of free 2d Yang-Mills in terms of branched holomorphic covers of the spacetime by the worldsheet \cite{Gross:1992tu, Gross:1993hu}. Here too, the proposals for the dual were topological string theories albeit of a somewhat non-traditional kind \cite{Horava:1993aq,Cordes:1994sd}. After all, the symmetric product 2d CFT is not too far removed from 2d Yang-Mills --- it arises as the infrared limit of 2d Super Yang-Mills-matter systems.  

We also see many features of the string theory which are expected of the tensionless limit. The fact that the correlators in the tensionless worldsheet theory get contributions localised to certain stringy saddle points in moduli space is also very much like what was seen by Gross and Mende in their classic analysis of high energy scattering in flat space string theory \cite{Gross:1987ar}. In fact, the logarithmic profile of the transverse direction $\Phi(z,\bar{z})$ near any of the insertions (\ref{coversol}) is very much like the Coulomb potential saddle points of \cite{Gross:1987ar}. In our case, the full string theory and not just the high energy sector is given by a saddle point reflecting the uniform nature of the tensionless limit in $\text{AdS}$.

Relatedly, we had already observed in Section 7 of  \cite{Eberhardt:2018ouy} that the enhanced higher spin symmetries \cite{Gross:1988ue,Gaberdiel:2014cha} of the free orbifold CFT is reflected in the free worldsheet theory of the $\mathrm{PSU}(1,1|2)_1$ WZW model. Here we have seen that even in the NSR formalism, the physical states of the $k=1$ theory can be described by a free worldsheet theory. (This observation was at the heart of the semiclassical exactness of the path integral.) This origin of higher spin symmetries in a free worldsheet suggests many directions for exploration, as also mentioned at the end of \cite{Eberhardt:2018ouy}. Moreover, the fact that we have a quasi-topological description also fits well with old ideas about topological strings describing a symmetry unbroken phase of string theory \cite{Witten:1988xj}. 
 
In particular, the Higgsing of these symmetries and their role in constraining the theory away from the tensionless limit is something we can try to concretely address with our new understanding of the relevant worldsheet correlators. Note that in the matching of the spectrum in \cite{Gaberdiel:2018rqv, Eberhardt:2018ouy}, we had already identified the moduli of the orbifold CFT with states in the worldsheet theory. When one deforms away from the free spacetime CFT, it is reasonable to expect that correlation functions in the worldsheet CFT will no longer be delta-function-localised but rather become smeared. It will be interesting to understand how this smearing occurs. In this context, the techniques being developed in \cite{Eberhardt:2018exh, Cho:2018nfn} for R-R deformations are likely to be useful. 

We should mention that while our Ward identity analysis of the correlators determines their structure, we have so far not yet solved them completely, see e.g.\ the function $W_\Gamma$ in eq.~(\ref{eq:localisation solution}) or the overall normalisation constant of the $3$-point function (\ref{eq:3point}).\footnote{The fact that the semiclassical path integral reduces to the Lunin-Mathur answer, which reproduces the right three point functions \cite{Lunin:2000yv, Lunin:2001pw}, suggests that this should work out correctly.} One would expect that these data will be  determined by null-vectors and crossing symmetry, and furthermore, that the constraints on the worldsheet will correspond precisely to those that are relevant for the dual CFT. Again, it would be interesting to work this out in more detail. More generally, given that we now have a much better control over correlators of spectrally flowed vertex operators for the $\mathrm{SL}(2,\mathds{R})$ WZW model, it should  be possible to understand the structure of this CFT much better. Ultimately, a proof of crossing symmetry generalising \cite{Teschner:2001gi} to spectrally flowed sectors should be given. 

We have discussed, in this paper, mostly the localisation for genus zero, but as we have already explained in Section~\ref{sec:genus}, we expect the basic structure to be the same at arbitrary genus. Our main argument for the localisation relied on the affine Ward identities on the worldsheet. These exist also on higher Riemann surfaces, but become significantly more complicated \cite{Eguchi:1986sb, Bernard:1988yv}. It would be interesting to see whether they put similar constraints on the correlation functions, which would give again a strong argument for localisation. Note that the semiclassical argument for covering map saddle points dominating the path integral goes through even for higher genus.  

We have concentrated on the case of ${\rm AdS}_3\times {\rm S}^3 \times {\cal M}_4$ with ${\cal M}_4 = \mathbb{T}^4$ or K3, but it would also be interesting to study the case ${\cal M}_4 = {\rm S}^3 \times {\rm S}^1$, for which the detailed worldsheet analysis was performed in \cite{Eberhardt:2019niq}, thereby confirming the duality proposal of \cite{Eberhardt:2017pty}. The worldsheet representations that appear in \cite{Eberhardt:2019niq} do not, in general, satisfy $j_i=\frac{k}{2}$, and it would therefore be interesting to see how the analysis of the correlators works in that case. 

To sum up, we now have at our disposal a concrete but yet highly nontrivial instance of the AdS/CFT correspondence where one has insight into how and why this amazing duality holds. It illustrates the fact that the tensionless limit of AdS string theories can serve as a useful starting point for deriving this correspondence. We indeed hope that this example will pave the way for a general derivation of gauge-string duality. 

\section*{Acknowledgements}

We thank Ofer Aharony, David Gross, Ben Hoare, Shota Komatsu, Juan Maldacena, Shiraz Minwalla, Shlomo Razamat, Suvrat Raju, Nati Seiberg, Ashoke Sen, Spenta Wadia and Edward Witten for useful discussions. All three of us are grateful for hospitality during the programme ``Higher Spins and Holography" at the Erwin Schr\"odinger Institute in Vienna. MRG and RG thank the organisers of the fifth Indo-Israeli workshop at Nazareth for providing the stimulating environment that facilitated the Immaculate Conception of this project. LE and MRG are grateful to CERN for hospitality during the workshop ``Exact Computations in AdS/CFT" at a late (but crucial) stage of this work. The work of LE and MRG was supported by the Swiss National Science Foundation through a personal grant and via the NCCR SwissMAP. LE gratefully acknowledges support by the Della Pietra Family at IAS. The work of RG is supported in part by the J. C. Bose Fellowship of the DST-SERB as well as in large measure by the framework of support for the basic sciences by the people of India.

\appendix

\section{The solution in the general case}\label{app:npointsol}

In this appendix, we essentially repeat the analysis of Section~\ref{sec:3pointproof} to show that \eqref{eq:localisation solution} is indeed a solution to the recursion relations of Sections~\ref{sec:recursion}.

To do so, we assume again that the insertion points are generic, so that infinity does not have to be treated as a special point. Moreover, since the relations are linear, we may prove them term by term in the sum of \eqref{eq:localisation solution}. Thus, in the following we shall fix one covering map $\Gamma$ corresponding to $w_1$, \dots, $w_n$ and show that
\be 
\Bigl\langle \prod_{i=1}^n V_{h_i}^{w_i}(x_i;z_i) \Bigr\rangle =\prod_{i=1}^n (a_i^{\Gamma})^{-h_i} \prod_{i=4}^n \delta(x_i-\Gamma(z_i)) \, W_\Gamma(z_1,\dots, z_n)\  \label{eq:localisation solution simplified}
\ee 
solves the recursion relations. We should mention that the function $W_\Gamma$ is further restricted by requiring the conformal Ward identities on the worldsheet. We also recall that $\Gamma$ in this formula denotes the covering map for which $\Gamma(z_4),\, \dots,\, \Gamma(z_n)$ are unspecified, see the discussion after \eqref{eq:localisation solution}.

Because of the delta functions in the correlators, we have the distributional identity
\be 
f(\Gamma(z_i))\, \Bigl\langle \prod_{i=1}^n V_{h_i}^{w_i}(x_i;z_i) \Bigr\rangle=f(x_i)\, \Bigl\langle \prod_{i=1}^n V_{h_i}^{w_i}(x_i;z_i) \Bigr\rangle
\ee
for any continuous function $f$. This means that one can effectively replace all $x_i$ by $\Gamma(z_i)$ in the formulae and vice versa. The situation is however more subtle than this since
\begin{align}
\hat{F}_0^j & = \partial_{x_j}\Bigl\langle \prod_{i=1}^n V_{h_i}^{w_i}(x_i;z_i) \Bigr\rangle \\
 & =  \prod_{i=1}^n (a_i^{\Gamma})^{-h_i}\,  \delta'(x_j-\Gamma(z_j)) \, 
\prod_{i\neq j }^n \delta(x_i-\Gamma(z_i)) \, W_\Gamma(z_1,\dots, z_n) +\cdots\ ,
\end{align}
where the further terms arise from derivatives acting on the $a_i$. The point to stress here is that the expressions will in general also contain (a single) first derivative of a delta function (in addition to the other $\delta$ functions). This single derivative can be turned into a usual $\delta$ function by means of the identity 
\be 
(x_j-\Gamma(z_j)) \, \delta'(x_j-\Gamma(z_j))=-\delta(x_j-\Gamma(z_j))\ .\label{eq:delta' distribution identity}
\ee
However, because of the appearance of these derivatives, we need to be very careful when performing the replacements $x_i \longleftrightarrow \Gamma(z_i)$. 

\subsection{The \texorpdfstring{$\delta'$}{deltap} terms} \label{app:deltap terms}

In order to deal with this issue, we shall first extract the $\delta'$ terms by defining 
\begin{align} 
\hat{G}^{(p)}(z)&=(x_p-\Gamma(z_p)) \, \hat{G}(z) \\
&=(x_p-\Gamma(z_p)) \sum_{i=1}^n\sum_{\ell=0}^{w_i} \frac{(\Gamma(z)-x_i)^2}{(z-z_i)^{\ell+1}} \hat{F}_\ell^i\ ,
\end{align}
where $p=4,\, \dots,\, n$, and $\hat{G}(z)$ is defined by 
\begin{align}
\hat{G}(z)=\sum_{i=1}^n\left(-\frac{2(\Gamma(z)-x_i)h_i}{z-z_i} \, \Bigl\langle \prod_{l=1}^{n} V^{w_l}_{h_l}(x_l;z_l) \Bigr\rangle +\sum_{\ell=0}^{w_i} \frac{(\Gamma(z)-x_i)^2}{(z-z_i)^{\ell+1}} \hat{F}_\ell^i\right)\ .\label{eq:G definition n point}
\end{align}
(This differs from the natural generalisation of the function $G(z)$, see \eqref{eq:G definition}, by multiplying through with the correlator.)  The prefactor $(x_p-\Gamma(z_p))$ guarantees that the only terms that contribute are those that contain the factor $\delta'(x_p-\Gamma(z_p))$. Multiplying by $(x_p-\Gamma(z_p))$ then turns $\delta'(x_p-\Gamma(z_p))$ into a regular $\delta$ function by means of the identity (\ref{eq:delta' distribution identity}), and thus $\hat{G}^{(p)}(z)$ only contains $\delta$ functions (and no derivatives of $\delta$ functions). Furthermore, all of them  necessarily arise from the second term in the bracket of $\hat{G}(z)$ --- the first term does not contain any derivatives of $\delta$ functions.

It therefore follows that we can directly perform the replacements $x_i \longleftrightarrow \Gamma(z_i)$ in the expressions of $\hat{G}^{(p)}(z)$. By the same reasoning as in Section~\ref{sec:3pointproof}, we have for  each $j=1,\ldots,n$ and each $m=0,\, \dots,\, w_j-2$
\begin{align} 
0 & = \partial_z^m \hat{G}^{(p)}(z_j)=\frac{(x_p-\Gamma(z_p))}{m!} \partial_{z}^m \sum_{i \ne j}  
\left. \sum_{\ell=0}^{w_i} \frac{(x_j-x_i)^2}{(z-z_i)^{\ell+1}} \hat{F}_\ell^i \, \right|_{z=z_j} \\
&=(x_p-\Gamma(z_p))\left. \left\langle\Bigl(J^-(z)-2x_j J^3(z)+x_j^2 J^+(z)\Bigr)\prod_{i=1}^n V_{h_i}^{w_i}(x_i;z_i) \right \rangle \right|_{(z-z_j)^m} \ ,
\end{align}
where we have dropped the first term in the expression for $\hat{G}(z)$ since it does not contribute to $\hat{G}^{(p)}(z)$ (as explained above).
The same argument as in Section~\ref{sec:3pointproof} shows now that
\be 
\hat{G}^{(p)}(z)= \alpha^{(p)} \partial \Gamma(z)\ ,
\ee
where $\alpha^{(p)}$ is again a function of $x_i$, $z_i$, $h_i$, $j_i$ and $k$. We can compute $\alpha^{(p)}$ by the same strategy as in the 3-point function case in Section~\ref{sec:3pointproof}, see eqs.~\eqref{eq:alpha derivation first}--\eqref{eq:alpha derivation last}. This leads to
\begin{align}
\alpha^{(p)} N=(x_p-\Gamma(z_p))\sum_{j=1}^n \left(a_j^\Gamma \hat{F}_{w_j}^j-h_j \Bigl\langle \prod_{i=1}^n V_{h_i}^{w_i}(x_i;z_i) \Bigr\rangle\right)=0\ ,
\end{align}
which vanishes since the term in brackets does not contain any derivatives of $\delta$ functions. This is, by assumption, the case for the correlation functions, and it also applies similarly to $\hat{F}_{w_j}^j$ since this also equals a correlation function, albeit one for which $h_j$ has been shifted to $h_j+1$, see  eq.~\eqref{eq:J+shift}.

Thus, we conclude that while the $\hat{F}^j_\ell$ themselves generically contain derivatives of $\delta$ functions, these cancel out once we consider $\hat{G}(z)$. Thus $\hat{G}(z)$ will only contain delta functions themselves.

\subsection{The \texorpdfstring{$\delta$}{delta} terms} \label{app:delta terms}

Let us now start again, and try to apply the arguments of Section~\ref{sec:3pointproof} to $\hat{G}(z)$. First we want to show that $\partial_z^m\hat{G}(z_j)=0$ for every $j=1,\ldots,n$ and $m=0,\dots,w_j-2$ is equivalent to the constraint equations. For $j=1,\,2,\,3$, this works exactly as for the $3$-point function case of Section~\ref{sec:3pointproof}. Let us therefore consider the expansion around $z_j$ with $j\geq 4$. For $m=0,\, \dots,\, w_j-1$ we get 
\begin{align}
\frac{\partial_z^m}{m!}\hat{G}(z=z_j)&=\frac{\partial_{z_j}^m}{m!}  \left\langle\Bigl(J^-(z)-2x_j J^3(z)+x_j^2 J^+(z)\Bigr)\prod_{i=1}^n V_{h_i}^{w_i}(x_i;z_i) \right \rangle \Bigg|_{z=z_j} \nonumber\\ 
&\qquad+\frac{\partial_z^m}{m!}\sum_{i=1}^n\sum_{\ell=0}^{w_i} \frac{2(\Gamma(z_j)-x_j)(\Gamma(z)-x_i)}{(z-z_i)^{\ell+1}} \hat{F}_\ell^i\, \Big|_{z=z_j}\nonumber\\
&\qquad+\delta_{m,w_j-1}\big(-2a_j^\Gamma h_j + (a_j^\Gamma)^2 \hat{F}^{j}_{w_j}\big)\ ,\label{eq:hatG derivatives}
\end{align}
where the last term arises as in the second line of eq.~(\ref{eq:recursion relation derivation 1}), while the middle line comes from the last term in 
(\ref{eq:G definition n point}) and requires more effort. To see how to get it, we first expand out $(\Gamma(z)-x_i)^2$ near $z=z_j$ as
\begin{align}\label{step0}
(\Gamma(z) - x_i)^2  & = \bigl((\Gamma(z) - x_j) + (x_j - x_i) \bigr)^2 \nonumber \\ 
&  = (\Gamma(z) - x_j)^2 + 2 (\Gamma(z) - x_j)(x_j - x_i)  + (x_j-x_i)^2 \ . 
\end{align}
The last term in (\ref{step0}) is what actually appears  in the correlator, see eq.~(\ref{eq:rewriting Jm constraint}) (and hence is already contained in the first line of (\ref{eq:hatG derivatives})). The middle term can be simplified as  
\begin{align}
& \sum_{i\neq j} \sum_{\ell=0}^{w_i} \frac{2 (\Gamma(z)-x_j)(x_j-x_i)}{(z-z_i)^{\ell+1}} \hat{F}_\ell^i  
\cong \sum_{i\neq j} \sum_{\ell=0}^{w_i} \frac{2 (\Gamma(z_j)-x_j)(x_j-x_i)}{(z-z_i)^{\ell+1}} \hat{F}_\ell^i  \nonumber \\ 
& \cong \sum_{i\neq j} \sum_{\ell=0}^{w_i} \frac{2 (\Gamma(z_j)-x_j)(\Gamma(z)-x_i)}{(z-z_i)^{\ell+1}} \hat{F}_\ell^i  \ , \label{res1}
\end{align}
where $\cong$ means up to terms of order $(z-z_j)^{w_j}$. Here we have, in the first step, replaced $\Gamma(z) \mapsto \Gamma(z_j) + a_j^\Gamma (z-z_j)^{w_j}$; in the resulting expression $\hat{F}_\ell^i$ is then multiplied by $(\Gamma(z_j)-x_j)$, and hence contains at most a $\delta$-function (but not a $\delta'$-function any more). Then we can replace in a second step $(x_j-x_i) \mapsto (\Gamma(z)-x_i)$. 

Finally, for the first term in (\ref{step0}) we write 
\begin{align} \label{step1}
(\Gamma(z) - x_j)^2 & = \Bigl( \bigl(\Gamma(z) - \Gamma(z_j)\bigr) + \bigl( \Gamma(z_j) - x_j \bigr) \Bigr)^2 \\
& =  \bigl(\Gamma(z) - \Gamma(z_j)\bigr)^2 + 2 \bigl(\Gamma(z) - \Gamma(z_j)\bigr) \cdot \bigl( \Gamma(z_j) - x_j \bigr) + 
\bigl( \Gamma(z_j) - x_j \bigr)^2 \ . \nonumber
\end{align}
The last term vanishes when multiplying $\hat{F}_\ell^i$, since $\hat{F}_\ell^i$ contains at most a $\delta'$-function, and the first term goes as $(a_j^\Gamma)^2 (z-z_j)^{2w}$, and hence only contributes if $i=j$ and $\ell=w_j-1$ --- in fact, this is precisely the very last term of eq.~(\ref{eq:hatG derivatives}). We therefore only have to deal with the middle term of (\ref{step1}) which leads to 
\begin{align}
& \sum_{i=1}^{n}\sum_{\ell=0}^{w_i} \frac{2 (\Gamma(z) - \Gamma(z_j)) (\Gamma(z_j) - x_j)}{(z-z_i)^{\ell+1}} \hat{F}_\ell^i  
\cong \sum_{\ell=0}^{w_j} \frac{2 (\Gamma(z) - \Gamma(z_j)) (\Gamma(z_j)-x_j) }{(z-z_j)^{\ell+1}} \hat{F}_\ell^j    \nonumber \\ 
& \cong \sum_{\ell=0}^{w_j} \frac{2 (\Gamma(z)-x_j) (\Gamma(z_j)-x_j) }{(z-z_j)^{\ell+1}} \hat{F}_\ell^j  \ , \label{res2}
\end{align}
where we have first used that only the term $i=j$ contributes (because of the factor of $(\Gamma(z) - \Gamma(z_j)$). In the resulting expression $\hat{F}^j_\ell$ is multiplied by $(\Gamma(z_j) - x_j)$, and hence only contains (at most) a $\delta$-function; thus we can replace in the other factor $\Gamma(z_j) \mapsto x_j$. Combining (\ref{res1}) with (\ref{res2}) then leads to the middle line of (\ref{eq:hatG derivatives}), i.e.\ to 
\be 
H(z)\equiv\sum_{i=1}^n\sum_{\ell=0}^{w_i} \frac{2(\Gamma(z_j)-x_j)(\Gamma(z)-x_i)}{(z-z_i)^{\ell+1}} \hat{F}_\ell^i\ .
\ee
Our next aim is to show that $H(z) =0$. To do so, we notice that $H(z)$ has no poles at the insertion points $z_i$. The residue of the potential single pole is absent since $\hat{F}^i_{w_i}$ does not contain derivatives of $\delta$ functions and hence vanishes upon multiplication with $(\Gamma(z_j)-x_j)$. $H(z)$ has potentially simple poles at $z_a^*$, the locations of the simple poles of $\Gamma(z)$. Their residue however vanishes, since
\be 
\mathop{\text{Res}}_{z=z_{a}^*} \, H(z)= -2 \mathop{\text{Res}}_{z=z_{a}^*} \, \frac{G^{(j)}(z)}{\Gamma(z)}=0\ ,
\ee
and we have used that $G^{(j)}(z)$ vanishes identically, as shown above. Since also $H(z)\to 0$ as $z \to \infty$, we conclude that also $H(z)=0$ identically. 
\medskip

Coming back to eq.~(\ref{eq:hatG derivatives}), it now follows that 
\be 
\partial_z^m \hat{G}(z=z_j)=0
\ee
for $m=0,\, \dots,\, w_j-2$, since then only the first line contributes, and it vanishes by virtue of the constraint equations of Section~\ref{sec:basic strategy}, see eq.~(\ref{eq:J-const}). Thus, by the same arguments as in Section~\ref{sec:3pointproof} we deduce that 
\be 
\hat{G}(z)=\alpha \, \partial \Gamma(z) \, \Bigl\langle \prod_{l=1}^{n} V^{w_l}_{h_l}(x_l;z_l) \Bigr\rangle \ .
\ee
Finally, we compute $\alpha$ along the lines of eqs.~\eqref{eq:alpha derivation first} -- \eqref{eq:alpha derivation last}. The first steps are identical, and we only show the changes from eq.~(\ref{penultimate}) onwards,
\begin{align}
\alpha N \Bigl\langle \prod_{l=1}^{n} V^{w_l}_{h_l}(x_l;z_l) \Bigr\rangle &=-\sum_{a=1}^N \oint_{z_a^*} \mathrm{d}z\ \frac{\hat{G}(z)}{\Gamma(z)} \nonumber \\
&=\sum_{j=1}^n \sum_{\ell=0}^{w_j}\left(\oint_{z_{j}} \mathrm{d}z\  \frac{\Gamma(z_j)+a_j^\Gamma(z-z_j)^{w_j}+\cdots}{(z-z_j)^{\ell+1}} \hat{F}_\ell^j-\delta_{\ell,0} \Gamma(\infty) \hat{F}_0^j\right) \nonumber \\
&=\sum_{j=1}^n \left(\Gamma(z_j) \hat{F}_0^j+a_j^\Gamma \hat{F}_{w_j}^j\right) \ .
\end{align}
To compute the first term further, we notice that because of \eqref{eq:delta' distribution identity} we have for $j=4,\, \dots,\, n$
\be 
\Gamma(z_j) \hat{F}_0^j=x_j \hat{F}_0^j+\Bigl\langle \prod_{i=1}^n V_{h_i}^{w_i}(x_i;z_i) \Bigr\rangle \ .
\ee
Then we can use the Ward identity and conclude
\be 
\alpha=\frac{1}{N} \Bigg[\sum_{j=1}^n a_j^\Gamma {F}_{w_j}^j+\left((n-3)-\sum_{j=1}^n h_j\right) \Bigg]\ .
\ee
Finally, we again compute the derivative of $\hat{G}$ at $z_j$ to order $w_j-1$, which will yield the recursion relations. For $\hat{G}(z)$, we have already done this in \eqref{eq:hatG derivatives}. We then compare this with the same derivative of $\alpha \partial \Gamma(z) \langle \cdots \rangle$. Equality of the two expressions are the recursion relations. It is now straightforward to check that they are solved by \eqref{eq:localisation solution}, provided that eq.~(\ref{eq:j condition}) holds. Note that the 
correction term $(n-3)$ relative to the ``naive" answer comes from the derivatives of the delta functions.

\section{The ``secret" representation}\label{app:secret}

In this appendix we want to explain that in the Wakimoto representation of the worldsheet theory, there exists a natural representation $\mathscr{S}$ that behaves as the vacuum representation with respect to $\mathfrak{sl}(2,\mathds{R})_{k+2}$, but is non-trivial with respect to the action of $\gamma(z)$ and $\partial \Phi(z)$. Abstractly, the relevant representation can be described by the short exact sequence (with respect to $\mathfrak{sl}(2,\mathds{R})_{k+2}$)
\be 
0 \longrightarrow\sigma^{-1}(\mathscr{D}^+_{j=\frac{k+2}{2}})\longrightarrow \mathscr{S} \longrightarrow\sigma^{-1}(\mathscr{D}^-_{j=\frac{k}{2}}) \longrightarrow 0\ .
\ee
We should mention that $\sigma^{-1}(\mathscr{D}^+_{j=\frac{k+2}{2}})$ is the vacuum representation of $\mathfrak{sl}(2,\mathds{R})_{k+2}$, and hence there is indeed a state $\ket{S} \in \mathscr{S}$ with the required properties. 
The representation $\mathscr{S}$ is indecomposable with respect to the $\mathfrak{sl}(2,\mathds{R})_{k+2}$ current algebra, but it is irreducible with respect to the full (Wakimoto) algebra.

The ``vacuum" state $\ket{S}=\ket{0}$ in $\mathscr{S}$ is part of a family of states $\ket{h}$, on which the spectrally flowed zero-modes of the currents and $\gamma_1$ act as\footnote{Note that since $\gamma(z)$ has conformal dimension zero, a pole in $\gamma(z)$ corresponds to a non-trivial action of $\gamma_1$.} 
\begin{subequations}
\begin{align}
\gamma_1 \ket{h}&=\ket{h-1}\ , \\
J^3_0 \ket{h}&=h \ket{h}\ , \\
J_{-1}^+ \ket{h}&=(h+k+2) \ket{h+1}\ , \\
J_1^- \ket{h}&=h \ket{h-1}\ .
\end{align}
\end{subequations}
Furthermore, we have $J^a_n \ket{S} = 0$ for all $n\geq 0$. 
The eigenvalue of $(\partial \Phi)_0$ on $\ket{S}$ is
\begin{align}
(\partial \Phi)_0 \ket{S}&=\frac{1}{k}(-J_0^3+(J^+\gamma)_0)\, \ket{S}\\
&=\frac{1}{k}J^+_{-1}\gamma_1 \, \ket{S} \\
&=\frac{k+1}{k} \, \ket{S}\ .
\end{align}
This is almost the value we would have expected (namely $1$). However, since $\partial \Phi$ has a background charge, the residues may receive shifts. The actual value of the residues can be explained from the anomalous transformation behaviour of $\partial \Phi$ under conformal transformations that is described in Appendix~\ref{app:anomalous partialPhi}.

\section{The structure of \texorpdfstring{$\boldsymbol{\partial \Phi}$}{dPhi}}\label{app:anomalous partialPhi}

In this appendix, we explain the transformation behaviour of $\partial \Phi$ which in particular determines its background charge. To start with we note that 
\be 
[L_m,\partial \Phi(z)]=-\frac{m(m+1)}{2k}z^{m-1}+(m+1)z^m\partial \Phi(z)+z^{m+1}\partial^2\Phi(z)\ ,
\ee
which follows from using the formula for the stress energy tensor in terms of Wakimoto fields. This can be integrated to the transformation formula
\be 
\widetilde{\partial\Phi}(z)=\big(\partial f(f^{-1}(z))\big)^{-1}\left(\partial \Phi(f^{-1}(z))+\frac{\partial^2f(f^{-1}(z))}{2k\,\partial f(f^{-1}(z))}\right)\ .
\ee
Hence, under the conformal transformation $z \to -\tfrac{1}{z}$, we find
\be 
\widetilde{\partial \Phi}(z)=\frac{1}{z^2}\partial \Phi\left(-\frac{1}{z}\right)+\frac{1}{k z} \ , 
\ee
and thus, in particular,
\be 
(\partial \Phi)_0^\dag=-(\partial \Phi)_0+\frac{1}{k}\ .
\ee
This implies the anomalous charge conservation \eqref{eq:anomalous charge conservation sphere}. As a consequence there is the charge $-\tfrac{1}{k}$ sitting at infinity in $z$-space.
\smallskip

For the computation of the background charge in the $\partial \Phi$ correlator we also need to understand the transformation behaviour of $\partial \Phi$ with respect to the spacetime Virasoro algebra. Using the explicit form of these generators \cite{Giveon:1998ns} 
\be 
\mathcal{L}_m=\oint \mathrm{d}z\ \left((1-m^2) (\gamma^m J^3)+\tfrac{1}{2}m(m-1)(\gamma^{m+1} J^+)+\tfrac{1}{2}m(m+1)(\gamma^{m-1} J^-)\right)(z) \ , 
\ee
we find 
\begin{subequations}
\begin{align} 
[\mathcal{L}_m,\partial \Phi(z)]&=\frac{1}{2}m(m+1) (\gamma^{m-1}\partial \gamma)(z)\ ,\\
[\mathcal{L}_m,\gamma(z)]&=-\gamma^{m+1}(z)\ .
\end{align}
\end{subequations}
Upon integration this leads to the (formal) transformation rule under a finite transformation 
\begin{subequations}
\begin{align} 
\widetilde{\partial \Phi}(z)&=\partial \Phi(z)-\frac{(\partial^2 f)(\gamma(z))}{2\,(\partial f)(\gamma(z))} \, \partial \gamma(z)\ , \\
\widetilde{\gamma}(z)&=f(\gamma(z))\ .
\end{align}
\end{subequations}
Thus under the conformal transformation $x \to -\tfrac{1}{x}$, we have
\be 
\widetilde{\partial \Phi}(z)=\partial \Phi(z)+ (\gamma^{-1}\partial \gamma)(z)=\partial \Phi(z)- (\widetilde{\gamma}^{-1}\partial \widetilde{\gamma})(z)\ .
\ee
Remembering that $\oint \mathrm{d}z\ \gamma^{-1}\partial \gamma=\mathcal{I}$ is the spacetime identity \cite{Giveon:1998ns, Eberhardt:2019qcl}, we see that the positions where $x$ is at infinity, $x=\infty$, carry an additional charge of $1$, thus correctly reproducing the structure we found in the main text.

\end{document}